\documentclass[twocolumn,preprintnumbers,amsmath,amssymb,aps,prb]{revtex4}
\usepackage{graphicx}
\begin{document}

\title{Dynamic Phases, Stratification, Laning, and Pattern Formation for Driven Bidisperse Disk Systems in the Presence of Quenched Disorder} 
\author{
 D. McDermott$^{1,2}$, 
Y. Yang$^{1,3}$,
C. J. O. Reichhardt$^{1}$, and  C. Reichhardt$^{1}$ 
} 
\affiliation{
$^1$Theoretical Division,
Los Alamos National Laboratory, Los Alamos, New Mexico 87545 USA\\ 
$^2$Department of Physics, Pacific University,
Forest Grove, Oregon 97116, USA\\
$^3$School of Physics and Astronomy, University of Minnesota,
Minneapolis, Minnesota 55455, USA\\
} 

\date{\today}
\begin{abstract}
Using numerical simulations,
we examine the dynamics of
driven two-dimensional bidisperse disks
flowing over quenched disorder.
The system exhibits a series of
distinct dynamical phases
as a function of applied driving force
and packing fraction
including a phase separated state as well as
a smectic state with
liquid like or polycrystalline features.
At low driving forces,
we find a clogged phase
with an isotropic density distribution,
while at intermediate driving forces
the disks separate into bands of high and low density
with
either
liquid like or
polycrystalline structure in
the high density bands.
In addition to the density phase separation,
we find that in some cases
there is a fractionation of the disk species,
particularly when the disk size ratio is large.
The species phase separated regimes
form a variety of patterns
such as large disks separated by chains
of smaller disks.
Our results show that the formation of laning states
can be enhanced by tuning the ratio of disk radius
of the two species,
due to the clumping of small disks
in the interstitial regions between the large disks.
\end{abstract}
\maketitle

\section{Introduction}
A large class of systems can be effectively described 
as a collection of interacting particles
moving over a random pinning landscape, where 
a variety of distinct dynamical phases appear
as a function of driving force \cite{1}.
Well studied examples of such systems include
vortices in type-II superconductors \cite{2,3,4,N1,N2},
driven Wigner crystals \cite{5,6}, 
skyrmions undergoing current-induced motion
\cite{8,9,10},
sliding pattern forming
assemblies coupled to random landscapes \cite{11,12}, 
colloids on disordered substrates \cite{13,14,15,16,17,18}, 
and active matter moving in complex environments \cite{19,20,21}.
These systems
often exhibit
multiple nonequilibrium phase transitions,
such as a transition from a pinned to a sliding phase
followed by transitions to different types of sliding phases.
Such transitions
are associated with clearly observable
changes in the velocity-force curves,
fluctuation spectra, and
spatial reordering of the particles.

Previous work on dynamical phase transitions in driven systems
has primarily focused on long or intermediate range
particle-particle interactions
that tend to favor a uniform
particle density,
such as that found
in magnetic or charged systems.
When particles of this type are placed on quenched disorder
composed of randomly placed strong pinning sites,
three nonequilibrium phases emerge:
a pinned disordered state, 
a plastic flow state
in which the particle positions are disordered and
the particles exchange neighbors as they move,
and a dynamically reordered anisotropic crystal or moving smectic
state that appears at high drives when the
effectiveness of the pinning is reduced.
\cite{1}.

There are numerous examples of systems in which the
particle-particle interactions are short ranged or steric,
including many types of colloidal suspensions,
emulsions, bubbles, 
and granular
matter.
Although it might be natural to assume that the short-range
interactions would produce simpler behavior than the longer-range
interactions when the particles are driven over quenched disorder,
it was recently shown
that monodisperse hard disks
moving over a random pinning landscape
exhibit a remarkably rich
variety of 
dynamical phases, 
including clogging, disordered plastic flow,
segregated flow, laning flow,
and
moving crystals \cite{22}.
The disk system can form moving density segregated states
containing high density bands coexisting with low density
regions.
In some cases, 
the dense bands form close packed hexagonal lattices
even when the overall density of the system is well below
the crystallization density.
At higher drives,
the crystalline bands
break up to 
form dense one-dimensional chains, while
at higher densities
the disks
form a moving crystalline solid \cite{22}.             
Density separated phases
cost no energy in systems with contact interactions,
since the
energy remains
small even when the particles accumulate in one region
and are depleted from another region.
In contrast,
when the interactions are longer range,
the system can minimize its energy by destabilizing
and
dispersing
any locally dense regions.

In this work,
we consider
bidisperse disks driven over quenched disorder
consisting of randomly placed pinning sites.
In the absence of driving or pinning, the disks
form a jammed solid at densities well below the crystallization
density $\phi=0.9$ of pin-free undriven monodisperse disks
\cite{23,24}.
Both monodisperse and bidisperse disks can
exhibit a density segregation into dense and depleted
regions, but the bidisperse disks can
also
undergo species segregation of the two disk sizes.
Numerous studies have demonstrated species
segregation
under nonequilibrium conditions
for short range repulsive bidisperse systems
including granular matter \cite{25,26,27,28}
and colloids \cite{29,30,31}, 
where the degree of segregation depends
on the ratio of particle sizes and the type of
driving force applied.
There are, however, few studies
examining the impact of quenched disorder
on size segregation.
An understanding of segregation effects in flowing
bidisperse disks coupled to quenched disorder not only
offers new insights on 
depinning and sliding phenomena,
but also could
be used to develop new methods for separating
or mixing
bidisperse or multidisperse systems of particles.
For example, some
geological systems 
can be described in terms of multidisperse disks
moving through random pinning,
and such systems could undergo
dynamic segregation.

This paper is organized as follows.  We describe our simulation technique
for the bidisperse disks driven over random pinning in Section II.
In Section~\ref{sec:1}, 
we show the dynamic patterns
that form for a system in which 50\% of the disks are large
and the radius ratio of the large to small disks is 1.4.
In Section~\ref{sec:2},
we consider large disks that are twice as big
as the smaller disks while maintaining the fraction of large
disks at 50\%.
In Section~\ref{sec:3},  
we show that by
reducing the fraction of large disks to 10\%,
we can enhance the
segregation and stratification effects.
We examine the scaling of the velocity-force curves
near depinning  
in Section~\ref{sec:4},
and we summarize our results in Section~\ref{sec:5}.

\section{Simulation}
\label{sec:simulation}
We consider a two dimensional (2D) system
of size $L \times L$
with periodic boundary conditions in the $x$ and $y$ directions.
The sample contains $N_{d}=N_s+N_l$ disks, where
$N_s$
disks have a small radius of $r_{s}$
and $N_l$
disks have a large radius of $r_{l}$. 
The disk dynamics are governed by the following overdamped equation of motion:
\begin{equation}
\eta \frac{d {\bf R}_{i}}{dt} = {\bf F}_{dd}  + {\bf F}_{p}  + {\bf F}_{D} .
\end{equation}
Here $\eta$ is the damping constant and 
${\bf R}_{i}$ is the location of disk $i$.
The   
disk-disk interaction force
is 
${\bf F}_{dd} = \sum_{i\neq j}k(r_{dd}^{ij} - R_{ij})\Theta(r_{dd}^{ij} - R_{ij}) {\hat {\bf R}_{ij}}$,
where
$r_{dd}^{ij}=r_i+r_j$,
$r_{i(j)}$ is the radius of disk $i(j)$, 
$R_{ij} = |{\bf R}_{i} - {\bf R}_{j}|$,
$\hat {\bf R}_{ij}  = ({\bf R}_{i}-{\bf R}_j)/R_{ij}$,
$\Theta$ is the Heaviside step function,
and the spring constant $k = 50$
is large enough to prevent the disks from overlapping by more than
1\% of their radii.
The pinning force ${\bf F}_p$
is produced by $N_p$
pinning sites
modeled as randomly placed non-overlapping parabolic
wells cut off at a radius of $r_p=r_s$
that can each capture at most one disk with a
maximum pinning force of $F_{p}=1.0$.
The density $\phi$ of the system is
given by the
area covered by the disks,
$\phi = \pi (N_s r^2_{s}+N_l r^2_{l})/L^2$,
where $L = 60$ and $r_{s} = 0.5$.
We vary $r_l$ and set the radius ratio
$\Psi = r_l/r_s$
to $\Psi = 1.4$ in Sec.~\ref{sec:1} and
$\Psi = 2.0$ in Sec.~\ref{sec:2}.
In a previous study of the jamming of bidisperse disks
using this model
with $\Psi=1.4$,
the jamming density in a pin free sample is
$\phi_{j} \approx 0.845$ \cite{32}.
We set $N_p=1440$, giving a fixed pinning site density of
$\phi_p = N_p \pi r_p^2/L^2=0.31$.
Previous studies have shown that
increasing $\phi_p$ does
not alter the
behavior, but only shifts
the driving forces at which the dynamical transitions occur \cite{22}.
The driving force ${\bf F}_D=F_D{\bf \hat{x}}$
is applied uniformly to all disks and is incremented in intervals of
$\Delta F_D=0.05$,
where we wait at least
$5 \times 10^7$ simulation time steps
between increments to ensure that the flow has reached a steady state.
On each drive increment, we measure 
the species-dependent
disk velocities,
$\langle V_x^s\rangle = N_d^{-1}\sum^{N_d}_{i=1}({\bf v}_i \cdot {\hat {\bf x}})\delta(r_i-r_s)$
and
$\langle V_x^l\rangle = N_d^{-1}\sum^{N_d}_{i=1}({\bf v}_i \cdot {\hat {\bf x}})\delta(r_i-r_l)$,
where ${\bf v}_{i}$ is the
instantaneous velocity of disk $i$.
We generate species-dependent histograms of $P(v_x)$, the
distribution of velocities $v_x$ of the individual disks in the direction
of applied drive, by first allowing the system to reach a steady state
and then sampling the velocities every $\Delta t = 5 \times 10^5$ simulation
time steps.  The corresponding $P(v_y)$ is Gaussian distributed about
$v_y=0$ since the motion of the disks perpendicular to the driving
force is unbiased.
We also characterize the dynamic phases and phase 
transitions using velocity-force curves,
the transverse root mean square displacements,
and other measures of the
particle spacing and density. 

\section{Minimally Phase Separating System with
  $N_l=N_d/2$}
\label{sec:1}
\begin{figure}
\includegraphics[width=3.5in]{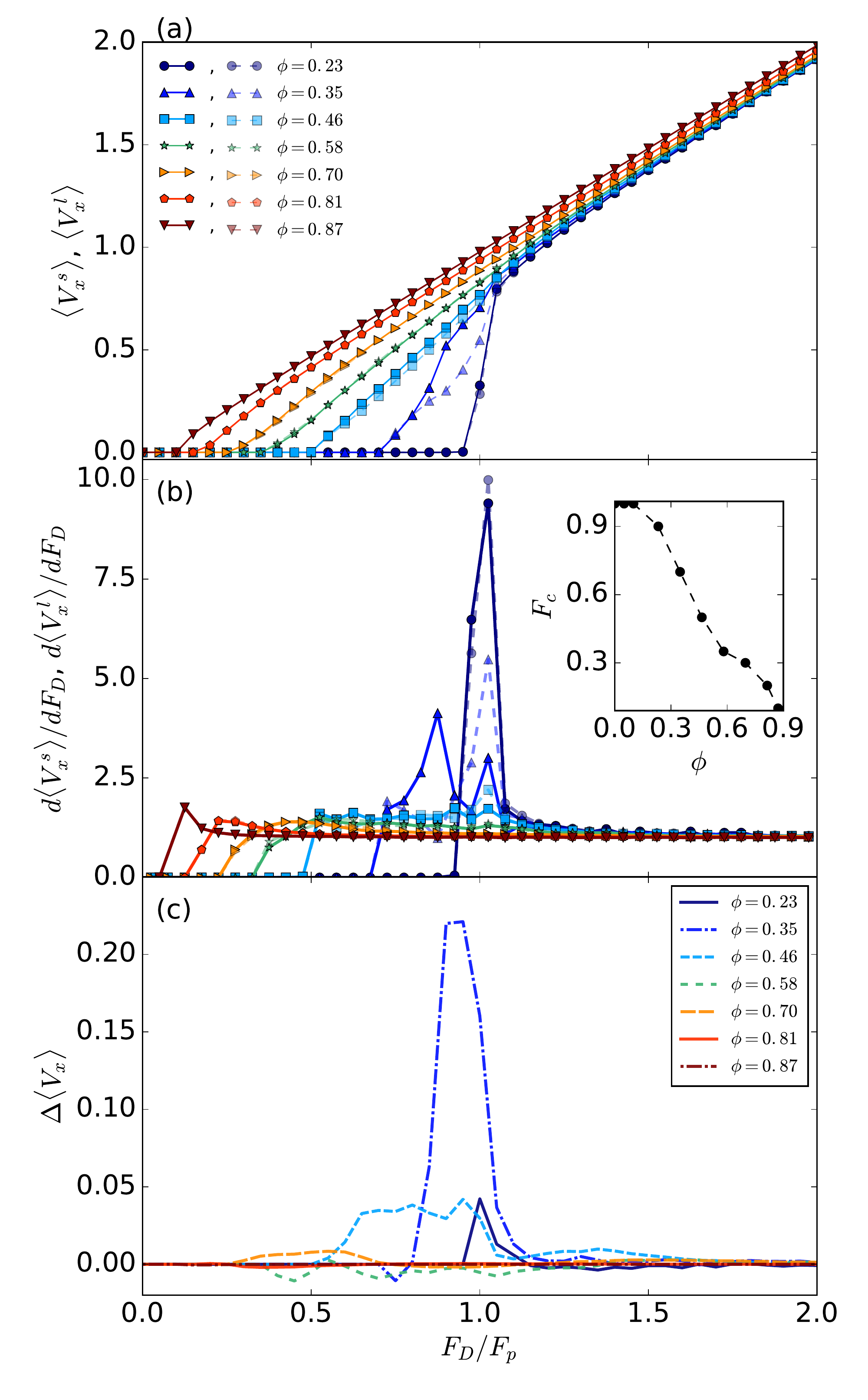}
\caption{
  (a) The species-dependent
  average disk velocities
  $\langle V_{x}^s \rangle$ (solid lines)
  and
$\langle V_{x}^l \rangle$ (dashed lines)
  versus driving force
  $F_{D}/F_{p}$
  in a sample with $\Psi=1.4$
  with equal numbers of small and large disks, $N_s=N_l$.
  The total 
  disk density
  is
  $\phi=0.87$ (down triangles), 
  $0.81$ (pentagons), 
  $0.70$ (right triangles), 
  $0.58$ (stars),   
  $0.46$ (squares), 
  $0.35$ (up triangles), 
  and 
  $0.23$ (circles). 
  (b) The corresponding $d\langle V_x^s\rangle/dF_D$ (solid lines)
  and $d\langle V_x^l\rangle/dF_D$ (dashed lines) vs $F_D/F_p$
  curves for the same values of $\phi$ showing a peak near
  $F_D/F_p=1.0$.
  Inset: critical depinning force $F_c$ vs 
  disk density $\phi$.
  (c) The difference
  $\Delta \langle V_x \rangle = \langle V_{x}^s\rangle - \langle V_{x}^l \rangle$
  vs $F_D/F_p$ for the same values of $\phi$ shown in panels (a) and (b).
}
\label{fig:1}
\end{figure}

We first consider samples with $N_s=N_l$ and
a disk diameter ratio of $\Psi = 1.4$. 
By varying the disk density from
$\phi=0.23$ to $\phi=0.81$,
we obtain a
ratio of pinning sites to disks in the range
$N_p/N_d = 2.0$ to $0.53$. 
At $\phi = 0.46$ there is one disk for every pin,
$N_p/N_d = 1.0$.
In Fig.~\ref{fig:1}(a) we
plot $\langle V_x^s\rangle$ and $\langle V_x^l\rangle$ versus
$F_D/F_p$
for $\phi=0.23$ to 0.87, and we
show the corresponding $d\langle V_x^s\rangle/dF_D$ and
$d\langle V_x^l\rangle/dF_D$ versus $F_D/F_p$ curves
in Fig.~\ref{fig:1}(b).
For $F_D/F_p \geq 1.5$, the velocities increase linearly
with drive
for all values of $\phi$.
In the
inset of Fig.~\ref{fig:1}(b)
we plot
the critical depinning force $F_c$ versus $\phi$. 
When $\phi$ is low,
$F_c \approx F_p$ since
each disk can be captured independently by a pinning site.
As the disk density increases, $F_c$ drops when the disks begin to interact
with each other.  Since each pin can capture at most one disk,
if an unpinned disk comes into contact with a pinned disk, the driving force
on both disks is offset by the pinning force on only one disk, lowering
the depinning threshold.  The number of disks in contact with each other
increases with increasing $\phi$, causing $F_c$ to decrease
monotonically.
We find no species dependence of $F_c$ at any value of $\phi$.
Figure~\ref{fig:1}(c) shows
$\Delta \langle V_x \rangle = \langle V_{x}^s\rangle - \langle V_{x}^l\rangle$,
the difference in net velocity between the two disk species.
This difference is largest in magnitude
near the depinning transition.

At a small disk density of $\phi=0.23$ in Fig.~\ref{fig:1}, both
$\langle V_x^s \rangle$ and $\langle V_x^l\rangle$ show relatively
sharp depinning transitions, as also indicated by the large
single peak at depinning
in the 
$d \langle V_x^s\rangle/dF_D$ and $d \langle V_x^l\rangle/dF_D$ versus
$F_D/F_p$ curves.
For drives close to but above $F_c$,
the smaller disks move slightly faster than the larger
disks so that $\Delta \langle V_x\rangle > 1$.

\subsection{Intermediate Disk Densities}

\begin{figure*}
  \includegraphics[width=0.75\textwidth]{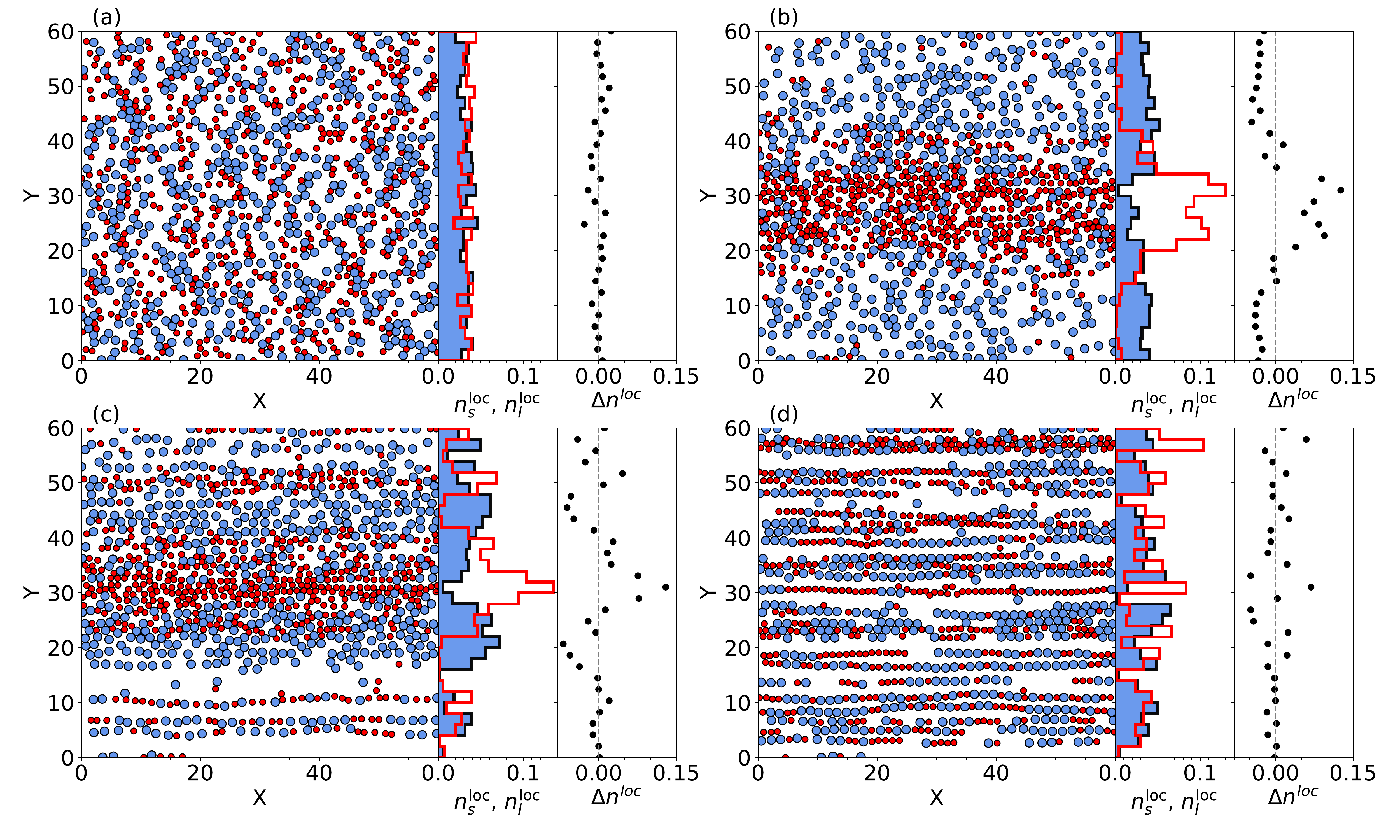}
  \hspace{-0.025\textwidth}
  \includegraphics[width=0.26\textwidth]{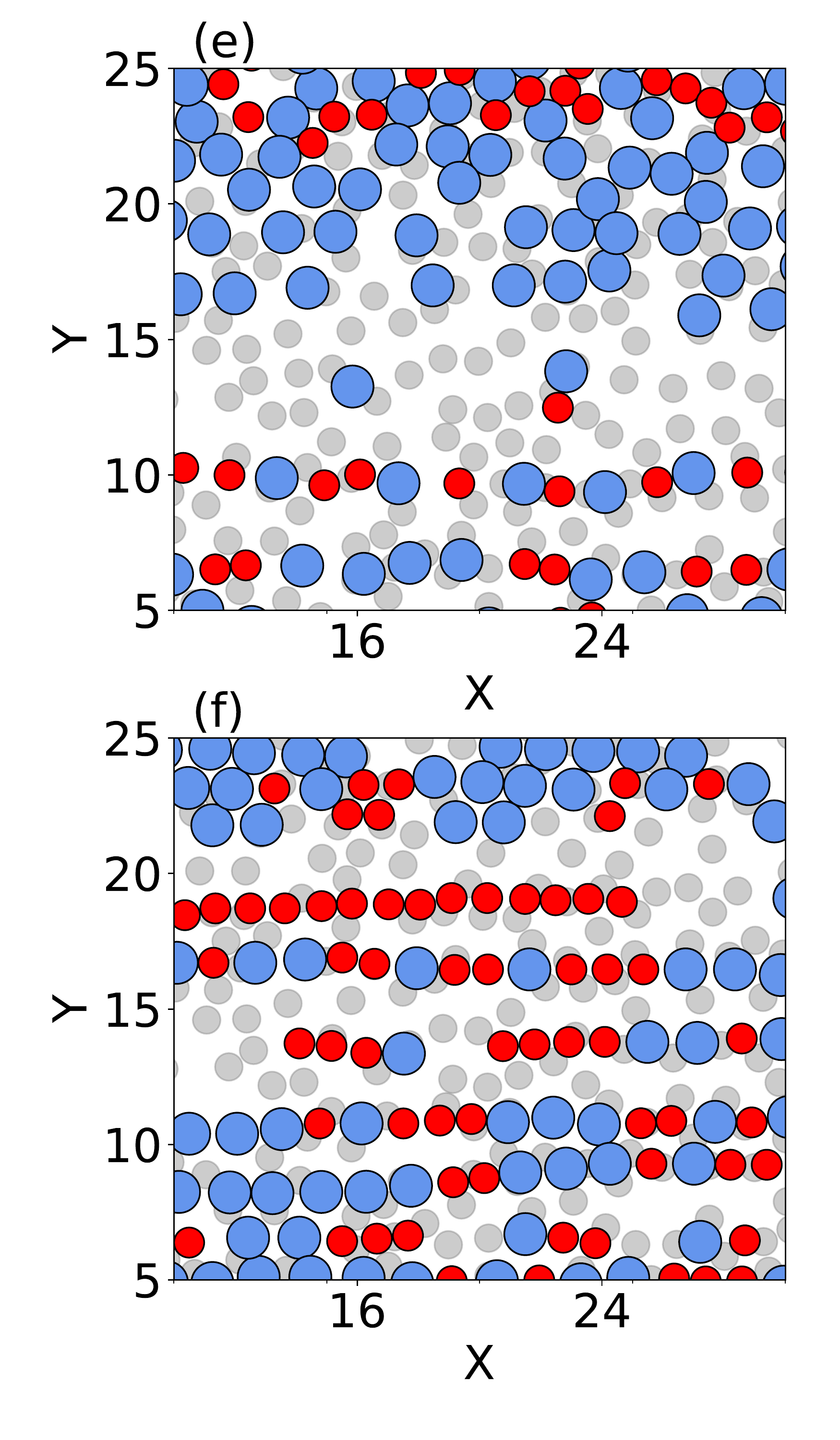}
  \caption{
    (a, b, c, d) Left panels: Large disk (blue circles) and small disk
    (red circles) positions  for the system
  in Fig.~\ref{fig:1} with $\Psi=1.4$ and $N_s=N_l$ at
  $\phi = 0.35$. 
  Center panels:
  $n_l^{\rm loc}$ (blue) and $n_s^{\rm loc}$ (red), the local number density of large and
  small disks, respectively,
  averaged over the $x$ direction for each $y$ position.
  Right panels:
  $\Delta n^{\rm loc}=n_s^{\rm loc}-n_l^{\rm loc}$
  plotted at each $y$ position.
  (a) The pinned state at $F_D/F_p=0.3$,
   where unpinned disks pile up behind pinned disks.
   (b) Just above depinning at $F_D/F_p=0.9$, where
   the sample contains a dense liquid-like region in the center
   surrounded by a gas-like region.
   (c) $F_D/F_p=1.1$, where the small and large disks become further segregated
   and the
   disks from the gas-like region collapse into chains with smectic ordering.
   (d) $F_D/F_p=2.0$, where the entire sample develops a smectic structure.
  (e) Detail showing large disk (blue circles), small disk (red circles),
  and pinning site (gray circles) locations in a portion of the sample
  from panel (c) at $F_D/F_p=1.1$.
  (f) Detail as in (e) for a portion of the sample from panel (d) at $F_D/F_p=2.0$.
  }
\label{fig:3}
\end{figure*}

Disk-disk interactions
become important at $\phi=0.35$,
where Fig.~\ref{fig:1}(b) shows that
a two peak structure emerges in
$d \langle V_x^s\rangle/dF_D$,
with one peak at $F_D/F_p=0.9$ and a smaller second peak at
$F_D/F_p=1.05$. We also find that $d \langle V_x^l\rangle/dF_D$ has
a small peak at $F_D/F_p=0.7$ and a larger peak at $F_D/F_p=1.05$.
A positive peak in $\Delta \langle V_x\rangle$
extends over the range
$0.8 < F_D/F_p < 1.05$
and is larger in magnitude than what we observe at other values of $\phi$.

In the left panel of Fig.~\ref{fig:3}(a) we illustrate the
disk
positions in the pinned state for $\phi=0.35$ at $F_D/F_p=0.3$.
Here, small numbers of unpinned disks have accumulated behind
pinned disks, giving a heterogeneous disk density and reducing
the depinning threshold to $F_c/F_p=0.7$.  In some regions,
short chains of disks composed preferentially of
large disks are stabilized at an angle to the driving direction.
In the center panel of Fig.~\ref{fig:3}(a)
we plot the local number density 
$n_l^{\rm loc}$ and $n_s^{\rm loc}$
of large and small disks, respectively, obtained by taking slices
of width $w=4r_s$ through
the sample at a fixed value of $y$ and dividing the number of disks of each
type in that slice by the slice area.
Thus,
$n_s^{\rm loc}(y)=(4r_sL)^{-1}\sum_{i}^{N_s} \Theta(|R_y^i-y|-2r_s)\delta(r_i-r_s)$
and $n_l^{\rm loc}(y)=(4r_sL)^{-1}\sum_{i}^{N_l} \Theta(|R_y^i-y|-2r_s)\delta(r_i-r_l)$.
The difference in local number density,
$\Delta n^{\rm loc}=n_l^{\rm loc}-n_s^{\rm loc}$,
is shown as a function of $y$ in the rightmost panel of Fig.~\ref{fig:3}(a).
Below the depinning transition, both disk species are distributed
uniformly throughout the sample.

\begin{figure}
  \includegraphics[width=0.48\textwidth]{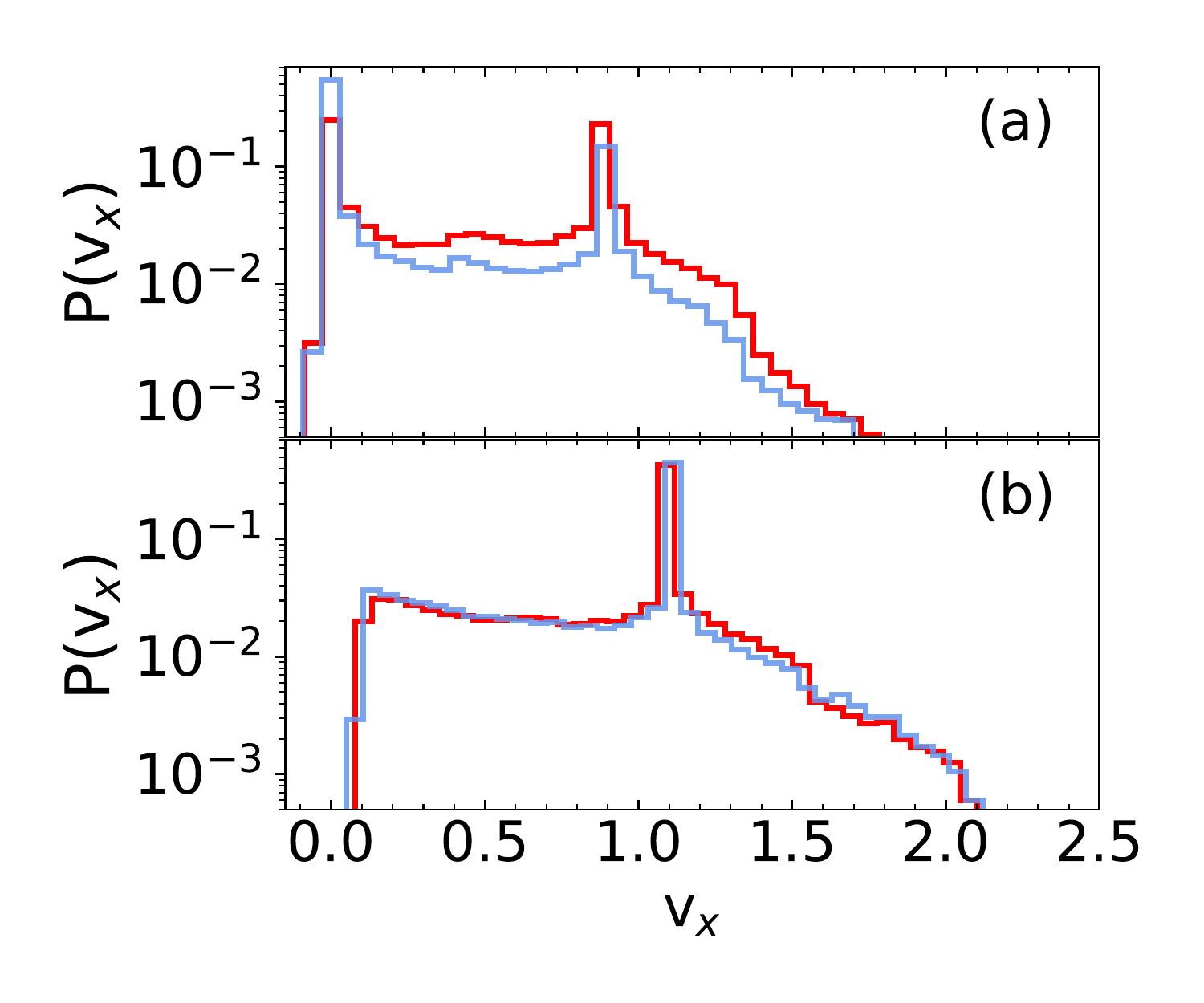}
  \caption{
    Histograms of $P(v_x)$ for the velocity $v_x$
    parallel to the driving
    direction 
    for the small disks (red) and large disks (blue)
    for the system in Fig.~\ref{fig:1}
    at $\phi = 0.35$
    with (a) $F_D/F_p =0.9$
    and (b) $F_D/F_p=1.1$.
}
\label{fig:5}
\end{figure}

Figure~\ref{fig:1}(c) shows that
for $\phi=0.35$ at $F_D/F_p = 0.9$,
the velocity of the small disks is larger than that of the
large disks,
giving
$\Delta \langle V_{x}\rangle \approx 0.24$.
At this drive
the sample develops
a horizontal band containing a high local density
of small disks
moving through a homogeneous distribution of large
disks, as illustrated in Fig.~\ref{fig:3}(b).
The peak in $d\langle V_x^s\rangle/dF_D$ at $F_D/F_p=0.9$ coincides with the
emergence of the dense band of small disks in the region
$10 < y < 45$.
At $y=30$ the value of $n_l^{\rm loc}$ is nearly zero, but in the
rest of the sample $n_l^{\rm loc}$ is roughly constant.
The small disks flow continuously while the large disks undergo
stick-slip motion that is enhanced in the vicinity of
the band of small disks, as shown in the supplementary
video \cite{supp3b}.
The species-dependent velocity distributions $P(v_x)$
in Fig.~\ref{fig:5}(a) show that
$v_x$ is bimodal for each species,
with peaks at
$v_x = 0$ and $v_x=0.9$ arising from the 
alternating pinned
and freely flowing motion of each disk.
The $v_x=0.9$ peak is higher for the small disks than
for the large disks since the small disks are more likely
to move freely
due to their separation into a dense band, and similarly
the peak at $v_x=0$ is highest for the large disks, which
are more likely to fall into a pinning site due to their greater radius.
Strong interactions with the pinning sites are required to
produce the $v_x=0$ peak.  Although $P(v_x)$ falls
off rapidly above $v_x=F_D=0.9$, there is still a tail with
finite weight at
$v_x>F_D$ produced by
disks that undergo brief rapid motion just after
escaping from a pinning site.

In Fig.~\ref{fig:3}(c)
at $F_D/F_p=1.1$,
the band of small disks in the
$\phi=0.35$ system
becomes more diffuse.
Simultaneously, the large disks
segregate into dense
bands surrounding the original band of small disks, while the lower density portion of
the sample develops smectic ordering consisting of chains of mixed disk sizes that
are oriented with the driving direction.
We find in Fig.~\ref{fig:1}(c) 
that $\langle V_x^s\rangle$ is slightly larger than $\langle V_x^l\rangle$ at
this drive since the higher density band of small disks is able to move more efficiently
over the pinning sites, as illustrated in the supplemental video~\cite{supp3c}.
Figure~\ref{fig:3}(e) shows
a more detailed plot of the disk positions along with the pinning site locations
in a portion of the sample from Fig.~\ref{fig:3}(c)
containing both the dense band of large disks and the smectic chains.
The disk species are not segregated within the chains, and since the
pinning force and driving force are nearly equal, the disks do not experience
much transverse displacement as they traverse the pinning sites.
In the smectic state, $P(v_x)$ has a single peak at $v_x=1.1$
with equal weight for both species,
as shown in Fig.~\ref{fig:5}(b).
Interactions of the disks
with the pins in the lower density portions of the sample
produce a broad plateau in $P(v_x)$ over the
range $0.1 < v_x < 1.1$.  Since $F_D > F_p$, the pinning sites can only
slow the disks but cannot trap them, so
there is no longer a peak
at $v_x=0$.

At higher drives for $\phi=0.35$,
the smectic ordering spreads throughout the entire sample,
as shown in Figs.~\ref{fig:3}(d)
at $F_D/F_p=2.0$.
The detailed view of the sample in Fig.~\ref{fig:3}(e) illustrates that
the long chains of disks have greater species separation and reduced
fluctuations in the $y$ direction compared to the chains
which form at lower $F_D$.
The dynamics of this state are illustrated in the supplemental movie~\cite{supp3d}.
Similar lane formation was observed for a low density of monodisperse
disks driven over quenched disorder \cite{22}, and is due in part to the fact that
strong density modulations incur no energy penalty in systems with short range
interactions.  Although on average $\Delta n^{\rm loc} \approx 0$, indicating that
the large scale species segregation found at lower drives is lost, we find that
individual chains can be preferentially composed of a single species of disk.
The velocity distributions $P(v_x)$ are similar to those shown in
Fig.~\ref{fig:5}(b) but have a sharper peak
at $v_x = F_D$.

The moving smectic state we observe differs from those
predicted by theory\cite{34,35} 
and observed in simulations \cite{N2,36,37}
and experiments \cite{N1} to occur in driven systems with quenched disorder 
such as vortices in type-II superconductors confined to two dimensions.
The short-range nature
of the
disk-disk interactions
permits the emergence of
extreme chaining behavior
in which the disks are nearly in contact along the driving direction but
are well-spaced in the transverse direction.
In contrast,
superconducting vortices
strongly repel one another at short distances,
and thus have a more even spacing in the directions parallel and transverse
to the drive.
Adjacent vortex rows in the smectic state contain dislocations that can
glide along the driving direction and permit the rows to slide past
one another.
For the disk system, adjacent rows are noninteracting and can move completely
independently of each other.

\begin{figure}
\includegraphics[width=3.5in]{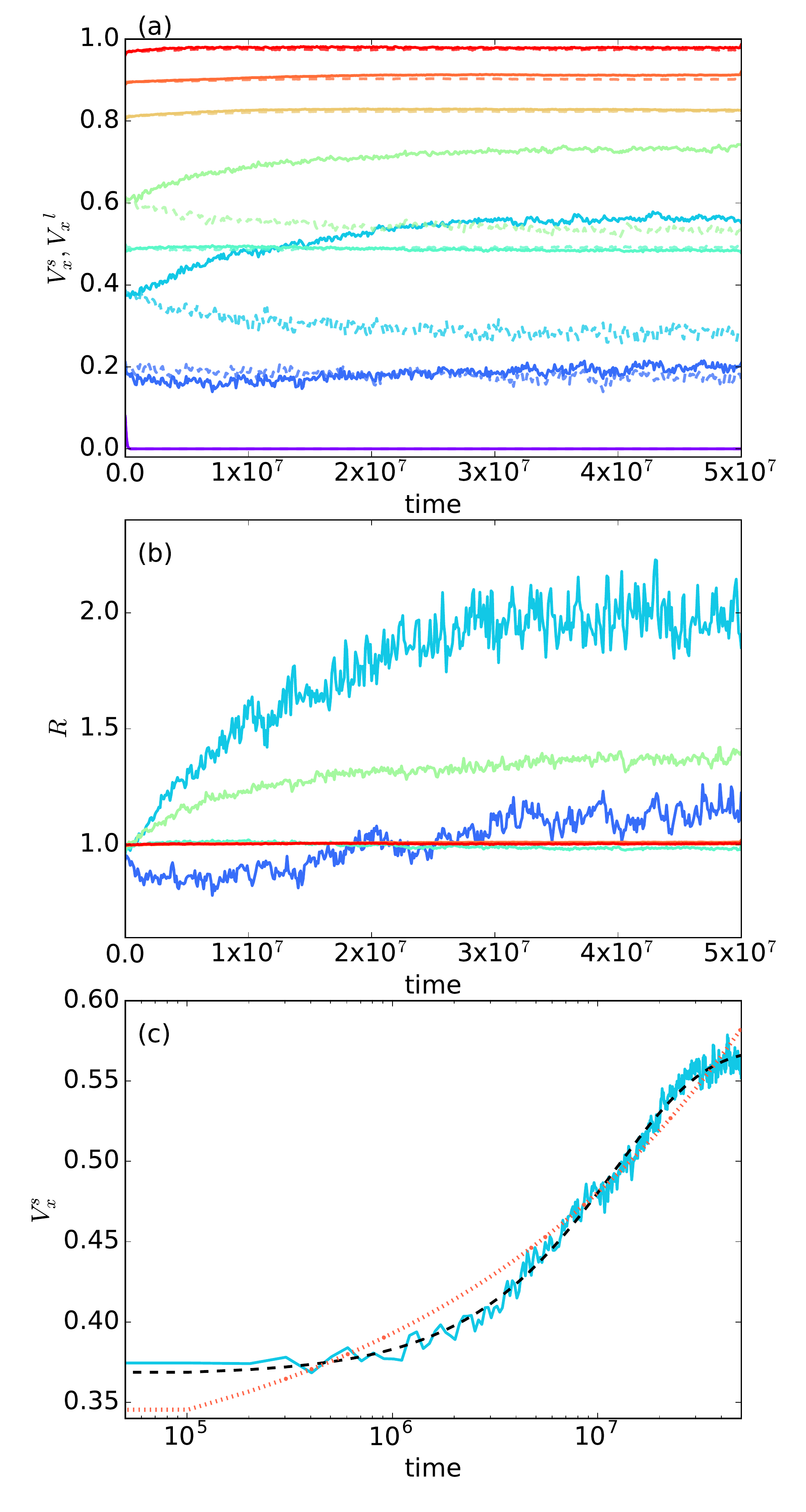}
\caption{
  (a) The instantaneous disk velocity 
  $V_{x}^s$ (solid lines) and $V_x^l$ (dashed lines)
  versus time 
  for the small and large disks, respectively,
  in the sample from
  Fig.~\ref{fig:1} at $\phi=0.35$ 
  for
  $F_D=0.7$, 0.8, 0.9, 0.95, 1.0, 1.05, 1.1, and 1.15, from bottom to top.
  (b) The corresponding ratio
  $R=V_{x}^s/V_{x}^l$ vs time for samples with $F_D/F_p>0.7$.
  (c) $V_{x}^s$ (solid blue line) vs time
  for the system in panels (a) and (b)
  at $F_D = 0.9$.
  Black dashed line: A fit to $V_x^s \propto e^{-t/\tau}$
  with $\tau=1.22 \times 10^7$.
  Red dot-dashed line: A fit to $V_x^s \propto t^{\alpha}$
  with $\alpha=0.26 \pm 0.01$.
}
\label{fig:2}
\end{figure}

In Fig.~\ref{fig:2}
we illustrate the time-dependent behavior of the $\phi=0.35$ system.
We find similar behavior when $0.2 < \phi < 0.5$.
Figure~\ref{fig:2}(a) shows
the instantaneous values of $V_x^s$
and $V_x^l$
versus time
at driving forces ranging from
$F_D/F_p = 0.70$ to  $1.15$.
In Fig.~\ref{fig:2}(b),
we show the corresponding ratio
$R = V_{x}^s/V_{x}^l$ versus time.
At $F_D \le 0.70$,
the disks are pinned,
and $V_x^s = V_x^l = 0$ except for a brief sharp decay at very early
times from a nonzero value.
At intermediate $F_D$ values of 0.75, 0.8, and 0.85,
we find large fluctuations in both $V_x^s$ and $V_x^l$, and
although the velocities of the two disk species are initially
identical, as the system evolves the velocities separate so that
at long times $V_x^s>V_x^l$.
At $F_D/F_p=0.9$, where the small disks first segregate into
a band, we can fit the velocity of the small disks to a stretched
exponential form, as shown in Fig.~\ref{fig:2}(c) where we find
$V_x^s \propto e^{-t/\tau}$ with $\tau=1.22 \times 10^7$.
For comparison, we show a fit to
$V_x^s \propto t^\alpha$ with $\alpha=0.26 \pm 0.01$, which gives
a poorer fit.
We find a similar stretched exponential behavior at
$F_D/F_p=0.95$,
and we show in Sec.~III.C that this behavior is associated with enhanced
transverse diffusion.
The stretched exponential time response suggests that the formation
of the segregated band of small disks is similar to an absorbing
phase transition of the type found in clogging systems \cite{Peter2017}.
For 
$F_D/F_p = 1.0$, 1.05, and $1.10$,
a stretched exponential fit gives a large time constant $\tau$, and
we show in Sec.~III.C that these drives produce superdiffusion in the
transverse direction.
At higher driving forces $F_D > 1.10$, the sample
quickly reaches a steady flow state with constant $V_x^s$ and $V_x^l$.

\subsection{High Disk Density}

\begin{figure*}
\includegraphics[width=\textwidth]{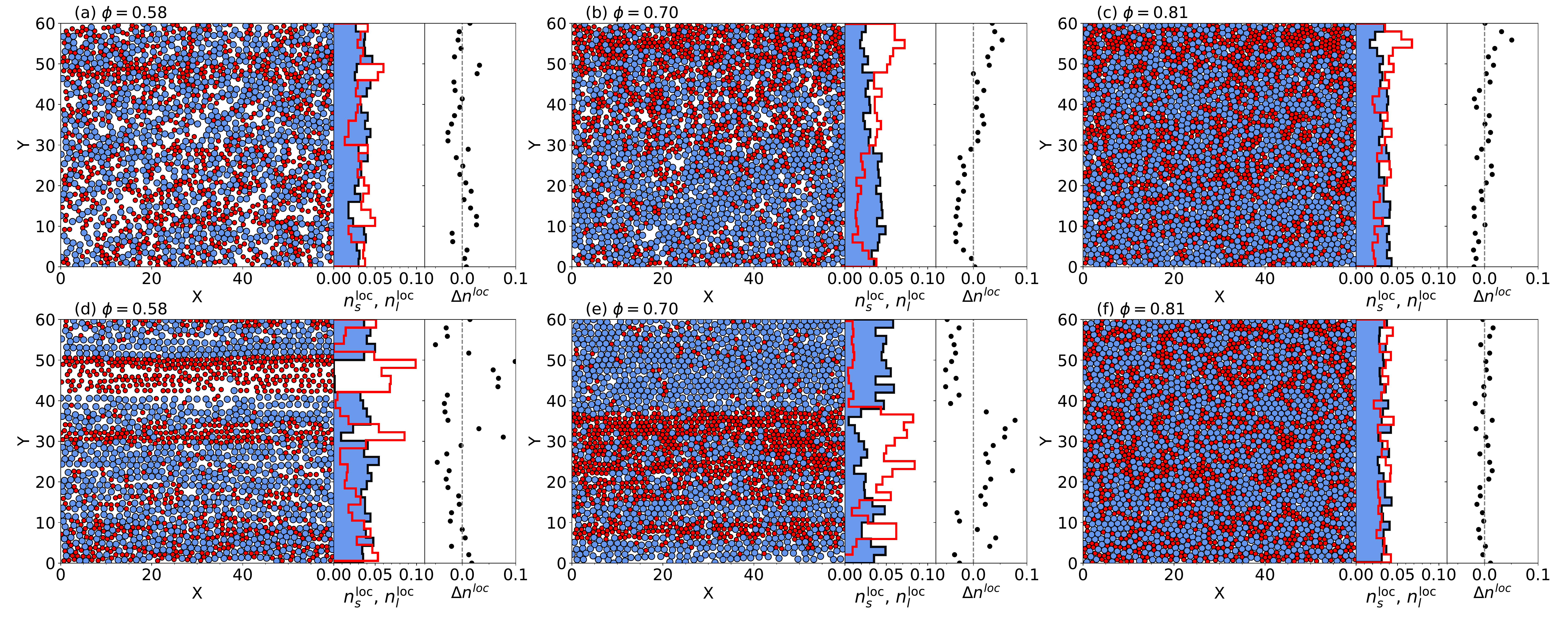}
\caption{
  Left panels: Large disk (blue circles) and small disk (red circles) positions for
  the system in Fig.~\ref{fig:1} with $\Psi=1.4$ and $N_s=N_l$.
  Center panels: $n_l^{\rm loc}$ (blue) and $n_s^{\rm loc}$ (red) as a function
  of $y$ position.  Right panels: $\Delta n^{\rm loc}$ as a function of $y$ position.
  (a) $\phi = 0.58$ and $F_D/F_p=0.5$, where there is a
  driven homogeneous phase.
    (b) $\phi = 0.70$ and $F_D/F_p=0.5$, showing 
  a segregated liquid.
    (c) $\phi = 0.81$ and $F_D/F_p=0.5$, where we find
    an isotropic polycrystalline phase.
    (d) $\phi=0.58$ and $F_D/F_p=1.3$, where
    the system fractionates into a liquid and smectic phase.
    (e) $\phi = 0.70$ and $F_D/F_p=1.3$, where
    the system is liquid throughout
    but forms 
    distinct horizontal bands.
    (f) $\phi = 0.81$ and $F_D/F_p=1.3$,
    which shows an isotropic polycrystalline state similar to that found
    at lower drives.
}
\label{fig:4}
\end{figure*}

\begin{figure*}
  \includegraphics[width=0.9\textwidth]{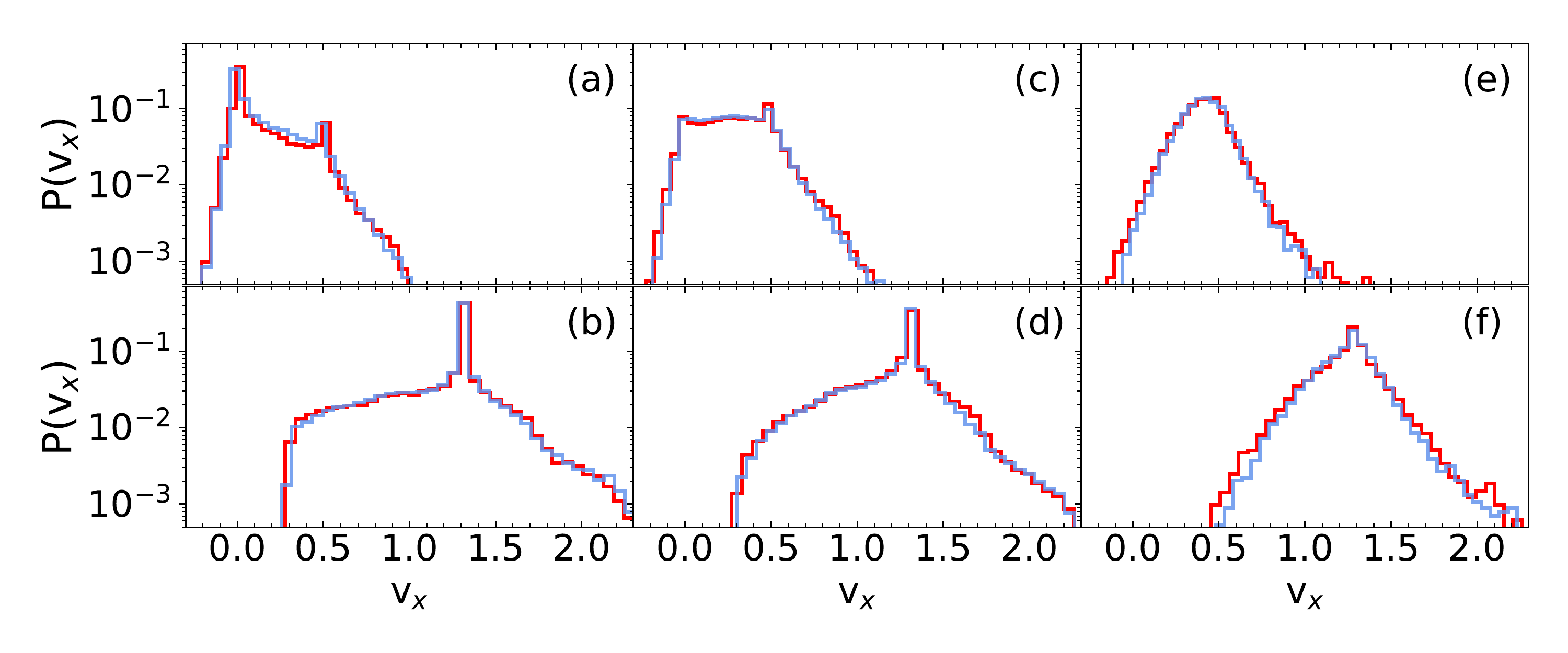}
  \caption{
    $P(v_x)$ for the small disks (red) and large disks (blue)
    for the system in Fig.~\ref{fig:1}
    at
    (a) $\phi=0.58$ and $F_D/F_p=0.5$;
    (b) $\phi=0.58$ and $F_D/F_p=1.3$;
    (c) $\phi=0.7$ and $F_D/F_p=0.5$;
    (d) $\phi=0.7$ and $F_D/F_p=1.3$;
    (e) $\phi=1.3$ and $F_D/F_p=0.5$;
    (f) $\phi=1.3$ and $F_D/F_p=1.3$.
}
\label{fig:6}
\end{figure*}

When $\phi = 0.46$,
the effect of interstitial or unpinned disks on the depinning
process becomes more important,
and the depinning threshold drops to
$F_c/F_p = 0.5$,
as shown in
Fig.~\ref{fig:1}.
The peak in $\langle V_x^s\rangle$ and $\langle V_x^l\rangle$ at depinning
is diminished in size, and we find that
$\Delta \langle V_{x}\rangle \approx 0.04$ 
over the range $0.5 < F_D/F_p < 1.0$.
At $F_D/F_p=0.5$, illustrated in Fig.~\ref{fig:4}(a),
$\Delta \langle V_x\rangle \approx 0$ and both types of
disks are in a gas-like state containing small regions of higher disk
density in the form of clumps and chains.  For this drive, the plots of
$n_l^{\rm loc}$ and $n_s^{\rm loc}$ in Fig.~\ref{fig:4}(a) show that
each disk species is uniformly distributed across the sample.
The corresponding velocity histogram $P(v_x)$ in Fig.~\ref{fig:6}(a)
shows a bimodal distribution produced by the stick-slip motion of
the disks, which are interacting strongly with the pinning sites.
The $v_x=0$ peak is higher than the $v_x=F_D$ peak, indicating that
the disks spend more time sticking and less time slipping, giving
a low value of $\langle V_x\rangle$ in Fig.~\ref{fig:1}(a).
At $F_D/F_p=1.3$ in Fig.~\ref{fig:4}(d),
where we again have $\Delta \langle V_x\rangle \approx 0$,
the disks phase segregate into
a liquid region surrounding a smectic region, which extends from
$40 < y < 55$.  The smectic state is characterized by strongly asymmetric spacing
of the disks, which are much closer together parallel to the drive than
perpendicular to the drive.  In this case, the smectic region
contains mostly
small disks and is of relatively low density.
The density of the liquid region varies as a function of $y$, and the liquid
is composed mainly of large disks separated by horizontal gaps for
$10 < y < 30$, while a densely packed liquid containing nearly equal numbers
of small and large disks appears for $y < 10$.
The large disks are almost completely depleted in the regions
$y \approx 30$ and $40<y<50$
but have a nearly uniform density in the rest of the sample,
as shown by the plot of $n_l^{\rm loc}$ in Fig.~\ref{fig:4}(d).
In Fig.~\ref{fig:6}(b), $P(v_x)$ has a single peak at $v_x=F_D=1.3$
and a broad distribution of velocities in the range $0.3 \leq v_x \leq 2.3$,
including a low velocity plateau.

For higher disk densities of
$\phi=0.58$ to 0.87,
$F_c$ continues to decrease with increasing $\phi$
while 
$\Delta \langle V_{x}\rangle$ becomes small.
The increased disk-disk interactions
that occur at the higher densities not only
diminish the depinning force, but also
equalize the velocities of each disk species due to the
higher frequency of disk-disk collisions.
In Fig.~\ref{fig:4}(b),
we show a $\phi=0.70$ sample
at $F_D/F_p = 0.5$,
where the disks are in a liquid state containing some small
localized clumps and chains.
There is some species segregation, with the small disks preferentially
located at the top of the sample and the large disks preferentially residing
in the bottom of the sample, as indicated by the plots of $n_l^{\rm loc}$ and
$n_s^{\rm loc}$ in Fig.~\ref{fig:4}(b).
We find a bimodal distribution of $P(v_x)$ as shown in
Fig.~\ref{fig:6}(c), but the two peaks are barely higher than the background
plateau
since the increased disk-disk interactions
reduce the effectiveness of the pinning sites.
The same sample at $F_D/F_p=1.3$ develops polycrystalline structure
in which the disks form wide species separated bands, as illustrated
in Fig.~\ref{fig:4}(e).
The polycrystalline clusters tend to be aligned in the driving direction.
Figure~\ref{fig:6}(d) shows a single peak in $P(v_x)$ at $v_x=F_D$ along with
a broad distribution of velocities over the range
$0.4 \leq v_x \leq 2.4$.  The plateau at low $v_x$ has vanished since
all of the disks are always moving at this drive, and it is replaced by a rapid
decrease in $P(v_x)$ with decreasing $v_x$.

At $\phi=0.81$, Fig.~\ref{fig:4}(c) shows that
when $F_D/F_p=0.5$,
the disks have a combination of liquidlike and polycrystalline structure.
Although the plot of $n_s^{\rm loc}$ indicates that there is a local increase
of small disk density near $y \approx 55$,
the disks are nearly jammed, and as a result further species segregation
is suppressed.
In Fig.~\ref{fig:6}(e), $P(v_x)$ has lost its distinct peaks and has a much
more Gaussian shape, since the strong interactions between the disks
prevent individual disks from being trapped by the pins.
At $F_D/F_p=1.3$ for the same sample in Fig.~\ref{fig:4}(f),
the disk structure is nearly the same except that any slight tendency
for segregation into a band has been destroyed.
The
plot of $P(v_x)$ in
Fig.~\ref{fig:6}(e)
shows a spread of velocities about $v_x=F_D$
due to the tightly packed motion of the disks.

For densities of $\phi=0.81$ and above,
the disks have a glassy arrangement at both low and high drives,
and the high packing fraction inhibits rearrangements of the
disks, preventing both species segregation and the realignment
of the polycrystalline regions with the driving direction.
We have tested the system for finite size effects using a
larger sample with $L=200$, where we found structures
similar to those illustrated
in Figs.~\ref{fig:3} and~\ref{fig:4}.
The only difference is that the large system can accommodate
multiple layers of segregated bands along the $y$ direction.

\subsection{Transverse Diffusion and Topological Order}
\label{sec:measures}
\begin{figure}
\includegraphics[width=3.5in]{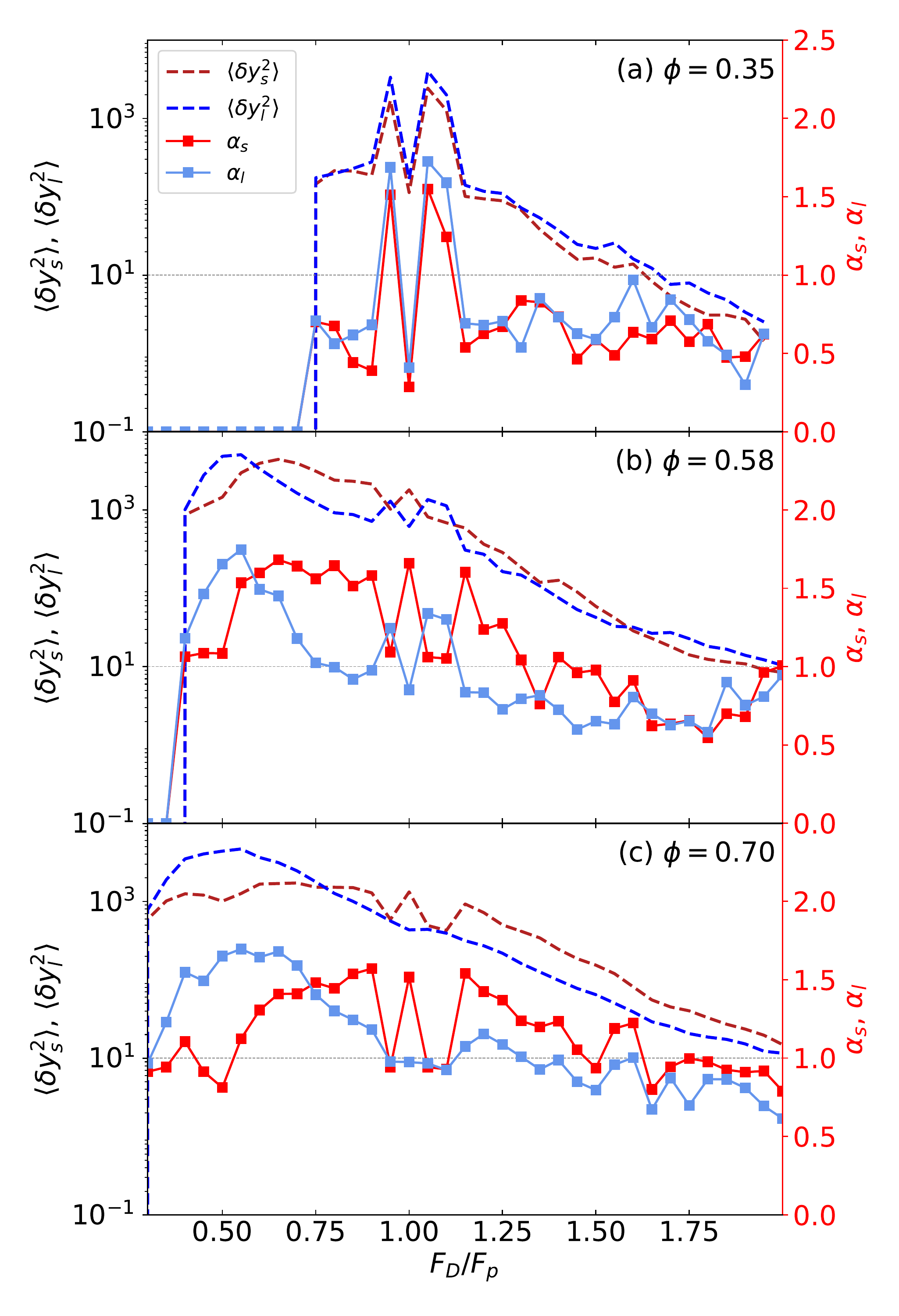}
\caption{
  The transverse displacements
  $\langle \delta y_s^2\rangle$ (red dashed line) and
  $\langle \delta y_l^2\rangle$ (blue dashed line) for the
  small and large disks
  obtained after $1\times 10^7$ simulation time steps
  vs $F_D/F_p$
  and the corresponding diffusive exponent $\alpha_s$ (red squares)
  and $\alpha_l$ (blue squares)
  for the system in Fig.~\ref{fig:1} at $\phi=$
  (a) $0.35$, 
  (b) $0.58$, and (c) $0.70$.
}
\label{fig:7}
\end{figure}

To further distinguish the phase behavior of each disk species, we
measure the disk displacements in the direction
transverse to the applied drive, 
\begin{equation}
  \langle \delta y_{s(l)}^2\rangle = \frac{1}{N_{s(l)}} \sum_{i=1}^{N_{s(l)}} [ y_i(t) - y_i(t_0)]^2,
\end{equation}
for the small and large disks, respectively.
In Fig.~\ref{fig:7} we plot
$\langle \delta y_s^2\rangle$ and $\langle \delta y_l^2\rangle$
obtained over the time interval $1\times 10^7$ to $5\times 10^7$
simulation time steps versus $F_D/F_p$
for samples with $\phi=0.35$, 0.58, and 0.7.
We also show
the corresponding 
diffusive exponents $\alpha_{s}$ and $\alpha_l$ obtained
from long-time fits to $\langle \delta y_{s(l)}^2\rangle \propto t^{\alpha_{s(l)}}$.
At all densities, 
$\langle \delta y_{s(l)}^2\rangle=0$ and $\alpha_{s(l)}=0$
for $F_D < F_c$
when the disks are motionless.
Previous studies of monodisperse disks showed superdiffusive transverse
flow
with $\alpha>1$ in regimes where density phase
separation occurred,
since the increased frequency of
disk-disk interactions in the high density region
produces a greater amount of disk motion transverse to the
driving direction \cite{22}.
The bidisperse disks
have a more complex behavior
since a wider variety of phase separated states occur that extend
down to lower densities.
In particular, the
large and small disks 
generally
exhibit
different transverse
diffusive behavior in the species separated regimes.

In Fig.~\ref{fig:7}(a) at
$\phi=0.35$,
both disk species
undergo subdiffusive transverse motion with $\alpha_{s(l)}<1$
when $F_D>F_c$.
Transverse movement is suppressed at low disk density
due to the infrequency of disk-disk collisions.
Near $F_D/F_p=1.0$, we find large fluctuations of
$\alpha_s$ and $\alpha_l$ due to the
gradual emergence of the
dense species separated bands
illustrated in Fig.~\ref{fig:3}(b,c).
At $F_D/F_p=0.9$ and 1.0, 
the dense liquid band of small disks is 
surrounded by a homogeneous low density
gas of large disks, and we find subdiffusive behavior
with $\alpha_{s(l)}<1.0$.
Superdiffusive behavior
with $\alpha_{s(l)}>1$ appears
at $F_D/F_p = 0.95$
where the small disks have more fully segregated into
a distinct horizontal band,
and also at $F_D/F_p=1.05$ and 1.1 where the small disks
form a smectic low density state containing horizontal chains.
Similar fluctuations
in $\alpha_{s(l)}$ appear near $F_D/F_p = 1$
for $0.35 < \phi < 0.5$, 
where 
some
samples reach a steady phase segregated,
particle separated state within $\Delta t = 5 \times 10^7$ time steps
while others do not.

In Fig.~\ref{fig:7}(b) at $\phi=0.58$
we find diffusive transverse motion with $\alpha_{s(l)} \approx 1$
whenever the disk density is homogeneous, including
near depinning
and
for driving forces at which
densely packed polycrystalline regions appear.
For drives just above depinning,
both types of disk undergo superdiffusive transverse motion as the
species separation illustrated
in Fig.~\ref{fig:4}(a) occurs.
The large disks transition to diffusive behavior
at $F_D/F_p = 0.75$, 
while the small disks remain superdiffusive 
until $F_D/F_p = 1.3$.
Above $F_D/F_p = 1.3$, the driving force
dominates the disk motion and the transverse
displacements are subdiffusive for both species.
In Fig.~\ref{fig:7}(c) at
$\phi=0.70$,
the
transverse motion is diffusive at depinning when
$F_D = F_c$.
The large disks are 
superdiffusive 
in the range $0.3<F_D/F_p<1.0$,
and
become diffusive at higher drives.
The small disks
are diffusive
for $0.3<F_D<0.5$,
superdiffusive
for $0.5<F_D<1.5$,
and diffusive above $F_D=1.5$.
A similar
intermediate superdiffusive phase
was observed
in Ref.~\cite{40}.
When the disk density is high, we find a transition from
diffusive to subdiffusive behavior coinciding with the
emergence of
a locked polycrystalline phase.
For example, at 
$\phi=0.814$, 
$\alpha_{s(l)} \approx 1$
for all $F_D>F_c$.
At $\phi=0.87$,
$\alpha_{s(l)}\approx 0$ since the disks are kinetically trapped.

\begin{figure}
\includegraphics[width=3.5in]{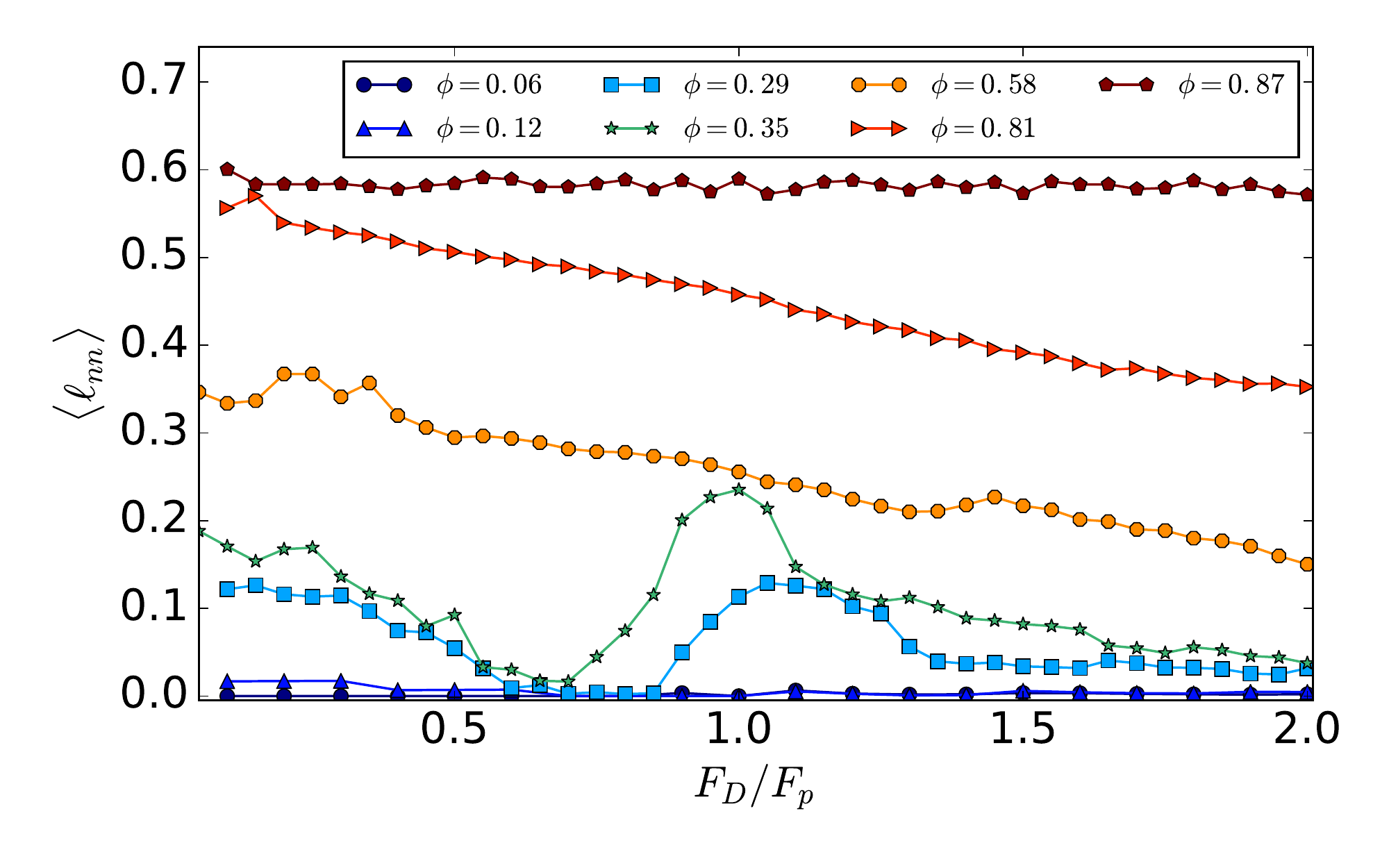}
\caption{ 
  The average transverse nearest neighbor distance
  $\ell_{nn}$ vs $F_D/F_p$
  for the system in Fig.~\ref{fig:1} at
  $\phi=0.06$ (blue circles),
  0.12 (blue triangles),
  0.29 (blue squares),
  0.35 (green stars),
  0.58 (orange circles),
  0.81 (red triangles),
  and 0.87 (brown pentagons).
}
\label{fig:8}
\end{figure}

To characterize lane formation,
we measure $\langle \ell_{nn}\rangle$, 
the average perpendicular distance between
disks that are in contact,
given by
\begin{equation}
    \langle \ell_{nn} \rangle = \sqrt{\langle \Theta(r^{ij}_{dd}-R_{ij})[{\bf R}_{ij}\cdot \hat{y}]^2 \rangle}, 
\end{equation}
where ${\bf R}_{ij}={\bf R}_i-{\bf R}_j$.
In Fig.~\ref{fig:8} we plot
$\langle \ell_{nn}\rangle$ versus $F_D/F_p$ for $\phi=0.06$ to $\phi=0.87$.
For small disk densities in the range
$\phi = 0.06$ to 0.12,
almost no disks are in contact with each other and
$\langle \ell_{nn} \rangle$ is nearly zero.
For higher disk densities,
in the pinned state
the disks tend to form blockages
perpendicular to the drive that
become more extensive as $\phi$ increases,
giving larger values of
$\langle \ell_{nn}\rangle$.
As the depinning threshold is approached, these blockages fall apart, so that
$\langle \ell_{nn} \rangle$
decreases monotonically over the range $0 < F_D < F_c$.
For
$\phi=0.29$
and $\phi=0.35$,
$\langle \ell_{nn} \rangle = 0$ just below depinning where
nearly all disk-disk contacts are lost, followed by a peak
in 
$\langle \ell_{nn} \rangle$
near 
$F_D/F_p = 1$,
where
phase segregation 
into
low and high density regions occurs.
When chain structures form at higher $F_D$,
$\langle \ell_{nn} \rangle$ plateaus to a small but finite value.
At $\phi=0.58$ and $\phi=0.81$,
$\langle \ell_{nn} \rangle$ decreases steadily for $F_D> F_c$,
where $F_c=0.4$ and 0.2, respectively.
At $\phi=0.87$, which is near the jamming limit,
$\langle \ell_{nn} \rangle \approx 0.6$ for all drives.

\begin{figure}
\includegraphics[width=3.5in]{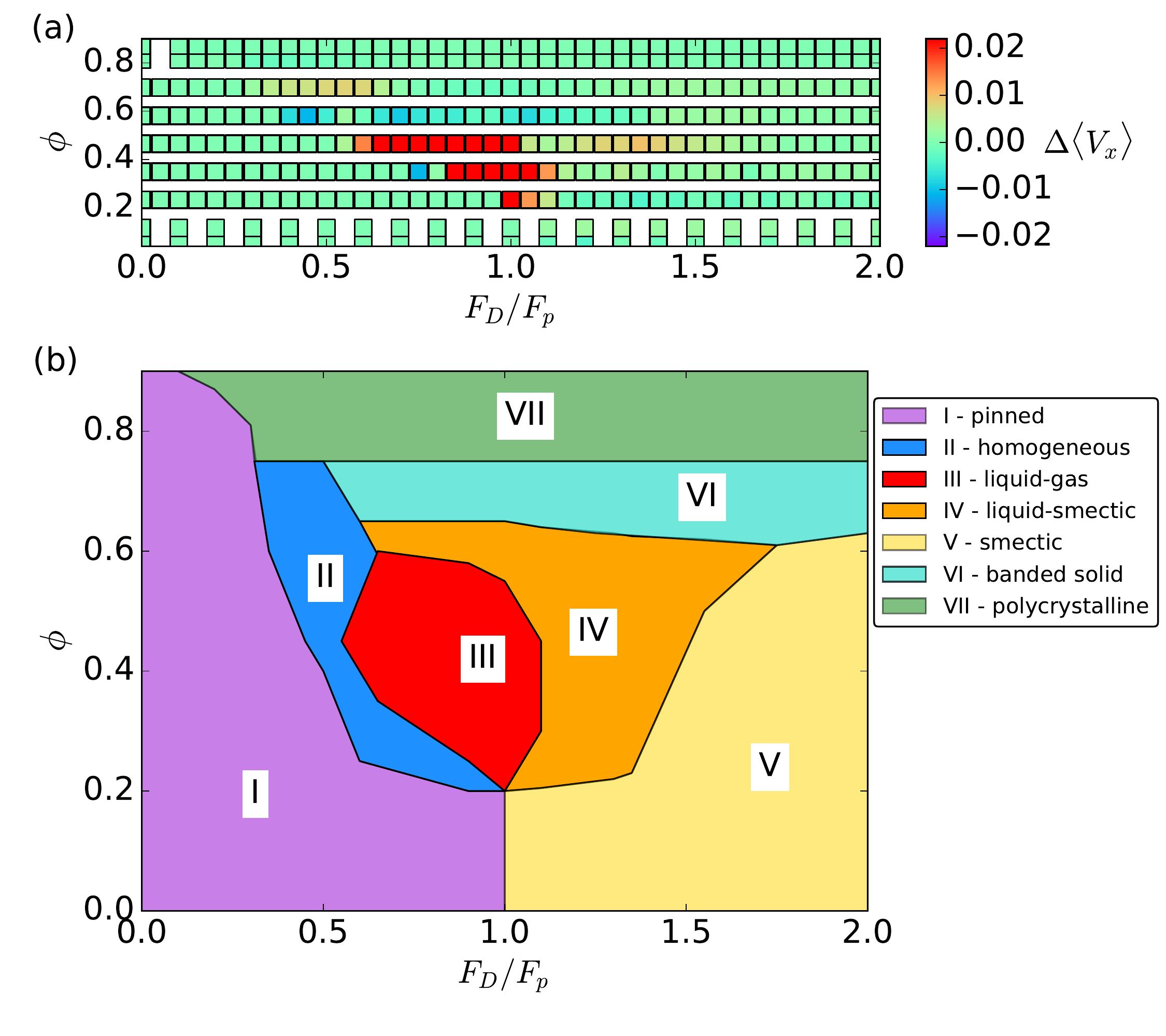}
\caption{
  (a) Heightfield plot of $\Delta \langle V_x\rangle$ as a function of
  total disk density $\phi$ vs driving force $F_D/F_p$, based on the data
  in Fig.~\ref{fig:1}(c).
  Red (blue) indicates that the velocity of the small disks
  is higher (lower) than that of the large disks.
  (b) A schematic dynamic phase diagram as a function of
  $\phi$ vs $F_D/F_p$.
  I: pinned or clogged;
  II: homogeneous flow; 
  III:  phase separated liquid-gas state;
  IV:  phase separated liquid-smectic, or moving chain, state;
  V: homogeneous smectic or moving chain state;
  VI: banded solid;
  VII: polycrystalline flowing state.
}
\label{fig:9}
\end{figure}

In Fig.~\ref{fig:9} we show a heightfield plot of the
$\Delta \langle V_x\rangle$ data from Fig.~\ref{fig:1}(c) as a
function of
disk density $\phi$
versus driving force $F_D/F_p$ 
for the $\Psi = 1.4$ system, while
in Fig.~\ref{fig:9}(b) we present a schematic dynamic
phase diagram as a function of $\phi$ vs $F_D/F_p$.
Phase I is the clogged or pinned state illustrated in
Fig.~\ref{fig:3}(a).  
Phase II, 
consisting of homogeneous plastic flow, is shown
in Fig.~\ref{fig:4}(a,b).
Phase III is the density phase separated liquid/gas state
from Fig.~\ref{fig:3}(b).
Phase IV, a density phase separated liquid/smectic state,
is illustrated in Figs.~\ref{fig:3}(c) and Fig.~\ref{fig:4}(d).
Phase V is the moving smectic/chain state
from Fig.~\ref{fig:3}(d).
Phase IV, the moving banded solid,
appears in Fig.~\ref{fig:4}(e), and
phase VII is the moving polycrystalline state
shown in Fig.~\ref{fig:4}(c) and (f).
Except for phase IV, we do not distinguish 
fractionation by species within the phases.
We note that the liquid-gas phase separation observed for monodisperse disks
in Ref.~\cite{22} is different in character from what we find here.
It occurs at higher disk densities of $\phi =0.46$ to $0.61$ 
and is associated with the formation of close-packed clusters of disks.

The boundary between the pinned phase I and the moving phases II, V, or VII
is determined by the
critical depinning force
plotted in Fig.~\ref{fig:1}(b).
At low $\phi$,
where the pins outnumber the disks, 
the system depins directly
into the moving smectic phase V.
As $\phi$ increases,
disk-disk interactions
become important and the homogeneous phase II flow
appears above depinning.
For intermediate $\phi$, this is followed at higher $F_D$ by density separation
into the liquid/gas phase III or the liquid/smectic phase IV,
while at higher drives the density becomes uniform again and the smectic
phase V emerges.
At higher $\phi$,
the disks are too dense to undergo phase separation and the system
transitions directly from the homogeneous phase II flow to
the banded solid phase VI.
For very large disk densities,
the disks can no longer exchange neighbors, and the system
depins into a moving polycrystalline phase VII.

\section{Enhanced Crystallization and Banding with Larger Radius Ratio at
  $N_l=N_d/2$}
\label{sec:2}

We next increase the radius ratio to $\Psi=2.0$, a value that is known to
produce phase separation for disks driven out of equilibrium \cite{25, 27}.
We fix $N_s=N_l$ and
consider disk densities in the range
$\phi = 0.19$  to $0.88$,
corresponding to
$N_D/N_p = 0.25$  to $1.125$. 
Here a disk density of
$\phi = 0.78$
corresponds to a ratio $N_p/N_D = 1.0$.

\begin{figure}
\includegraphics[width=3.5in]{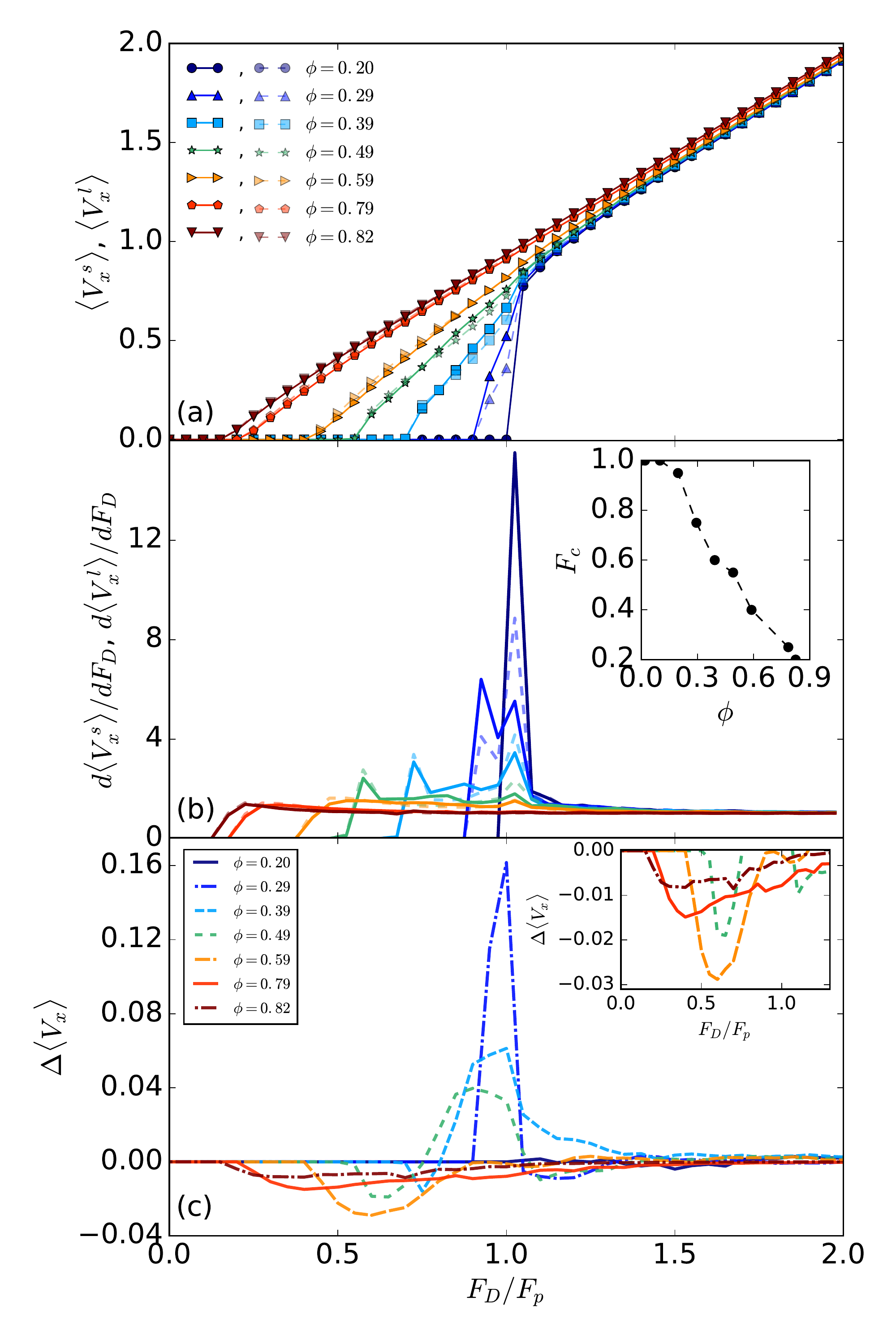}
\caption{   (a) $\langle V_{x}^s\rangle$ (solid lines)
  and $\langle V_x^l\rangle$ (dashed lines)
  vs
  $F_{D}/F_{p}$
  in a sample with $\Psi=2.0$ and 
  $N_s=N_l$ at
  $\phi=0.82$ (down triangles),
  $\phi=0.79$ (pentagons),
  $\phi=0.59$ (right triangles),
  $\phi=0.49$ (stars),
  $\phi=0.39$ (squares),
  $\phi=0.29$ (up triangles),
  and
  $\phi=0.20$ (circles).
  (b) The corresponding $d\langle V_x^s\rangle/dF_D$
  (solid lines) and $d\langle V_x^l\rangle/dF_D$ (dashed lines) vs $F_D/F_p$
  curves for the same values of $\phi$
  showing a peak near
  $F_D/F_p=1.0$.
  Inset: $F_c$ vs $\phi$.
  (c) The corresponding 
  $\Delta \langle V_{x}\rangle$
  vs $F_D/F_p$.
  Inset: a detail from the main panel of the region around $F_D/F_p=0.5$ where
  $\Delta \langle V_x\rangle < 0$ for large $\phi$.
}
\label{fig:10}
\end{figure}

The plot of $\langle V_x^s\rangle$ and $\langle V_x^l\rangle$ versus
$F_D/F_p$
in Fig.~\ref{fig:10}(a)
for the $\Psi=2.0$ system at different values of $\phi$ 
has similar behavior to that shown in Fig.~\ref{fig:1}(a),
with a pinned state at low drives,
a non-linear velocity-force relation
above depinning,
and a linear dependence of velocity on drive for high $F_D$.
The corresponding $d\langle V_x^s\rangle/dF_D$ and $d\langle V_x^l\rangle/dF_D$
versus $F_D/F_p$ curves in 
Fig.~\ref{fig:10}(b),
as well as the plot of $F_c$ versus $F_D/F_p$ in the inset
of Fig.~\ref{fig:10}(b), are also similar to what was shown
in Fig.~\ref{fig:1}(b).
In Fig.~\ref{fig:10}(c),
the plot of
$\Delta \langle V_x\rangle$ versus $F_D/F_p$ indicates
a higher velocity of the small disks at low $\phi$ similar to that found
in Fig.~\ref{fig:1}(c); however,
at low driving forces and high $\phi$,
we find that the large disks have a higher velocity than the small disks,
as highlighted in the inset of Fig.~\ref{fig:10}(c).

At the lowest density of $\phi=0.20$
in Fig.~\ref{fig:10}, the small and large disks both have the same
behavior, and
the depinning occurs sharply at $F_D/F_p = 1.0$,
with a distinct transition from pinned to elastic flow.
Since this system contains
fewer disks than
the $\Psi=1.4, \phi=0.23$ system,
the depinning transition is sharper, and the peak in
$d \langle V_{x}^s \rangle /d F_D$ and $d\langle V_x^l\rangle/ dF_D$
at $F_D/F_p=1.0$ is
larger.

\begin{figure*}
\includegraphics[width=\textwidth]{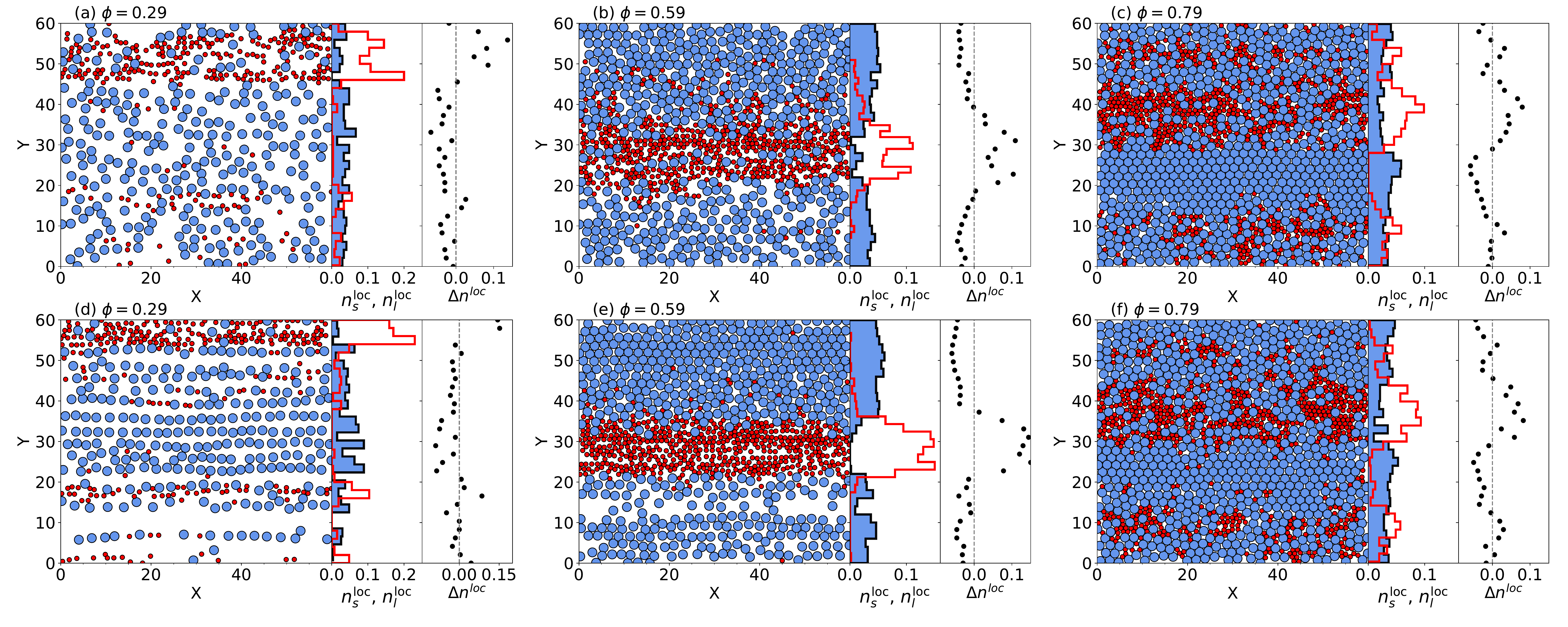}
\caption{
  Left panels: Large disk (blue circles) and small disk (red circles) positions
  for the system in Fig.~\ref{fig:10} with $\Psi=2.0$
  and $N_s=N_l$.  Center panels: $n_l^{\rm loc}$ (blue) and $n_s^{\rm loc}$
  (red) as a function of $y$ position.  Right panels: $\Delta n^{\rm loc}$ as a
  function of $y$ position.
  (a) $\phi=0.29$ and $F_D/F_p=0.95$.
  (b) $\phi=0.59$ and $F_D/F_p=0.9$.
  (c) $\phi=0.79$ and $F_D/F_p=0.9$.
  (d) $\phi=0.29$ and $F_D/F_p=1.1$.
  (e) $\phi=0.59$ and $F_D/F_p=1.1$.
  (f) $\phi=0.79$ and $F_D/F_p=1.1$.
}
\label{fig:11}
\end{figure*}

At $\phi=0.29$,
we find
an enhancement
in the velocity of the small disks near depinning
since the large disks can
easily be pinned by traps and other large disks,  
while the small disks slip through smaller apertures
to form a segregated dense band, as illustrated
in Fig.~\ref{fig:11}(a) at $F_D/F_p=F_c=0.95$.
Here the large disks are uniformly distributed through the sample,
while the small disks are concentrated in a band extending
from $45 < y < 60$.
This is the same type of segregation found in
Fig.~\ref{fig:3}(b).  
In Fig.~\ref{fig:10}(b),
$d \langle V_{x}^s \rangle /d F_D$ peaks at $F_D/F_p = 0.95$,
whereas
$d \langle V_{x}^l \rangle /d F_D$ peaks at $F_D/F_p = 1.0$,
indicating that the smaller disks begin to flow freely at lower drives
than the larger disks.
Above $F_D/F_p = 1$, there is a transition
to a liquid of small disks surrounded by a smectic state of large disks,
as illustrated in Fig.~\ref{fig:11}(d) for $F_D/F_p=1.1$.
This is accompanied by a large positive
peak in
$\Delta \langle V_x\rangle$
over the range
$1.05 < F_D/F_p < 1.25$,
as shown in Fig.~\ref{fig:10}(c).
The smectic chain structure of the large disks
increases
the number of disk-disk interactions
and
diminishes the effectiveness of the pinning for the large disks.
The small disks tend to form chains at higher drives.

Near depinning at
$\phi = 0.39$,
we find a density phase segregated state
containing
distinct bands of high density liquid smectic regions
and low density regions, similar
to the structure illustrated in Fig.~\ref{fig:3}(b) and (e).
There are two distinct peaks in
$d\langle V_x^s\rangle/dF_D$ and $d\langle V_x^l\rangle/dF_D$ in
Fig.~\ref{fig:10}(b)
near $F_D \approx F_c = 0.75$ where the small disks begin to move freely
and $F_D \approx F_p$ where the motion of the large disks increases.
In Fig.~\ref{fig:10}(c), $\Delta \langle V_x\rangle > 0$  over the range
$0.8 < F_D/F_P < 1.0$, indicating that the small disks can
flow more easily in the liquid smectic region, which they preferentially occupy.
At $\phi = 0.49$,
there is a pronounced crossover
in $\Delta \langle V_x \rangle$ in Fig.~\ref{fig:10}(c)
from
a negative value
for $0.6 < F_D/F_p < 0.7$
to
a positive value
for $0.8 < F_D/F_p < 1.0$,
indicating that the large disks are moving faster than the small
disks at lower drives but slower at higher drives.

For $\phi = 0.59$,
$\Delta \langle V_x\rangle$ is never positive but has an enhanced
negative region at low drives above depinning in the range
$0.4 < F_D/F_p < 0.9$,
as highlighted in the inset of Fig.~\ref{fig:10}(c).
Species segregation of the disks occurs in the window
$0.8 < F_D/F_p < 0.9$.
As illustrated in Fig.~\ref{fig:11}(b) for $F_D/F_p=0.9$,
the large disks form a cluster
that spans nearly the entire system,
while the small disks are concentrated in a band ranging from $20 < y < 40$.
The small disks
form
relatively few disk-disk contacts, making them less likely
to be depinned due to disk-disk interactions, and thus reducing
their velocity compared to the large disks.
At higher drives, all of the disks depin and the difference in velocity among
the two disk species drops to zero.
At $F_D=1.1$, shown in Fig.~\ref{fig:11}(e), 
the small disks remain in a single high density band
while the large disks form
a low density smectic state
at $0<y<10$ coexisting with
a high density liquid state containing
polycrystalline regions
at $35<y<60$.
A low density void region appears at
$10<y<20$.
The motion of the particles in this state is illustrated in
the Supplementary Material \cite{supp11b}.

When $\phi \ge 0.59$,
$d\langle V_x^s \rangle/F_D$ and $d\langle V_x^l \rangle/F_D$ have a
smooth rather than sharp increase
at $F_D = F_c$.
There is an extended regime in which the
velocity of the large disks is higher than that of the small disks,
with $\Delta \langle V_x\rangle < 0$
over the range $0.2 < F_D/F_p < 1.5$ for the $\phi=0.79$ system.
As shown in Fig.~\ref{fig:11}(c) for $\phi=0.79$ at $F_D/F_p=0.9$,
a significant fraction
of the large disks form tight polycrystalline packings
while 
the small disks 
form trapped clusters over specific horizontal windows.
The structure remains similar at higher drives, as shown
in Fig.~\ref{fig:11}(e) at $F_D/F_p=1.1$.
For larger systems with $L=200$ at high $\phi$, we find multiple large
polycrystalline regions rather than a single band spanning the system.

\subsection{Transverse Diffusion and Topological Order}

\begin{figure}
\includegraphics[width=3.5in]{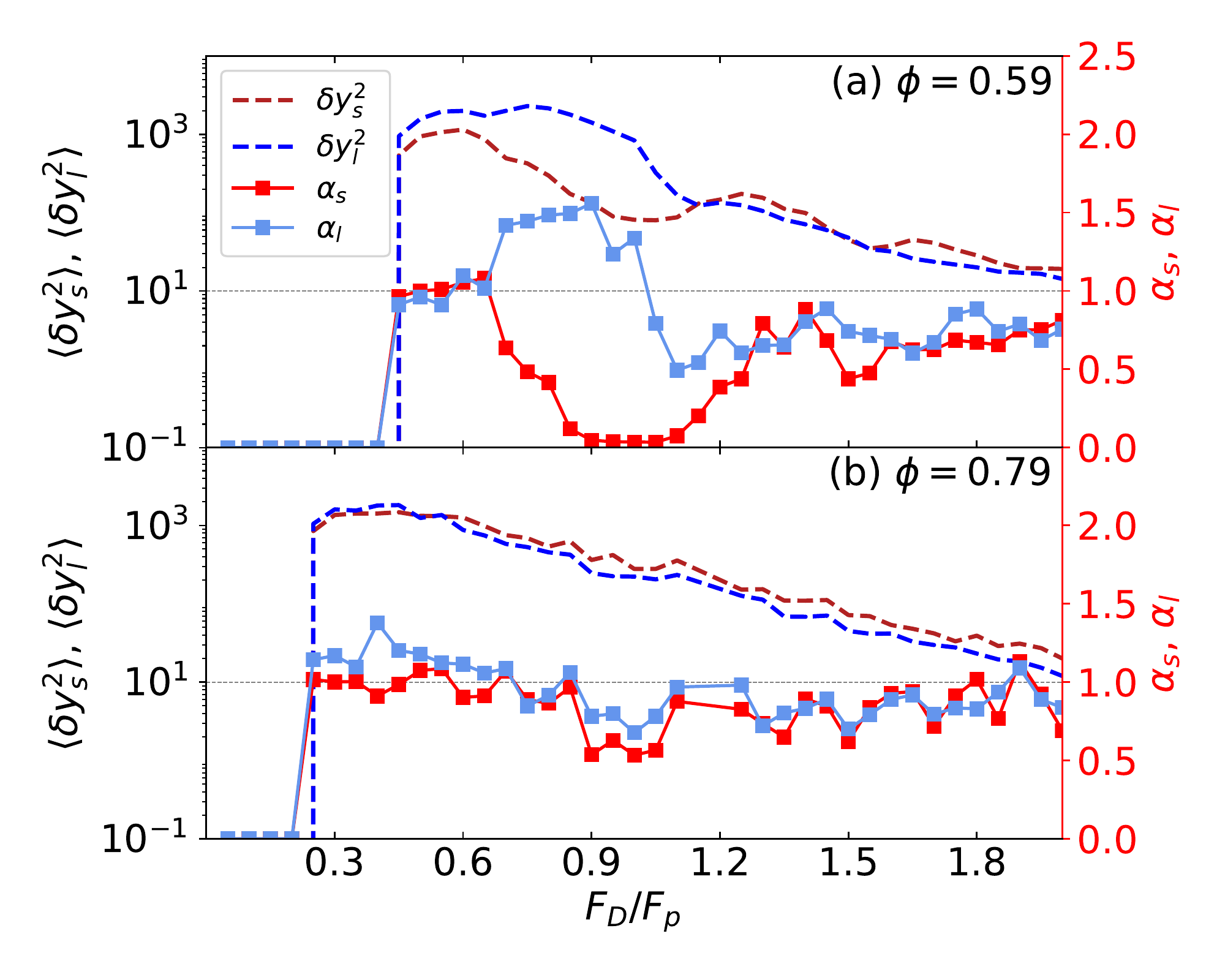}
\caption{ Transverse displacements $\langle \delta y_s^2\rangle$ (red dashed
  line) and $\langle \delta y_l^2\rangle$ (blue dashed line) for the small
  and large disks obtained after $1 \times 10^7$ simulation time steps
  vs $F_D/F_p$ and the corresponding diffusive exponent $\alpha_s$ (red
  squares) and $\alpha_l$ (blue squares) for the system in Fig.~\ref{fig:10} with
  $\Psi = 2.0$
  for densities $\phi = $ (a) $0.59$ and (b) $0.79$.
}
\label{fig:12}
\end{figure}

In Fig.~\ref{fig:12}, we plot
the transverse diffusion $\langle \delta y_s^2 \rangle$
and $\langle \delta y_l^2\rangle$ along with the exponents
$\alpha_s$ and $\alpha_l$ versus $F_D/F_p$
for the $\Psi=2.0$ system from Fig.~\ref{fig:10}.
At $\phi=0.59$ in Fig.~\ref{fig:12}(a),
we find homogeneous flow with $\alpha_s \approx \alpha_l \approx 1$
at low driving forces of
$0.4 < F_D/F_p < 0.6$,
indicating diffusive behavior.
At intermediate driving forces,
$0.6 < F_D/F_p < 1.2$,
the small disks are subdiffusive 
since they have become confined in a horizontal band,
as shown in Fig.~\ref{fig:11}(b) and (d).
The large disks are superdiffusive 
for $0.6 < F_D/F_p < 1.0$,
and
become subdiffusive at higher drives once their structure
changes
from a homogeneous liquid with small voids
to a denser liquid containing
a large horizontal gap.
The small voids permit a transverse flow of the large disks
that is suppressed once a large void opens at higher drives.
For $1.2 < F_D/F_p < 2.0$,
the driving force dominates the behavior of
both disk species, which form chain states that
move subdiffusively in the transverse direction.
At $\phi=0.79$ 
in Fig.~\ref{fig:12}(b),
$\alpha_s \approx \alpha_l  \approx 1$ at all driving forces above
depinning,
indicating diffusive transverse flow for both disk species.
This is expected in 
a liquid phase containing polycrystalline regions
of homogeneous density.

\begin{figure}
  \includegraphics[width=0.45\textwidth]{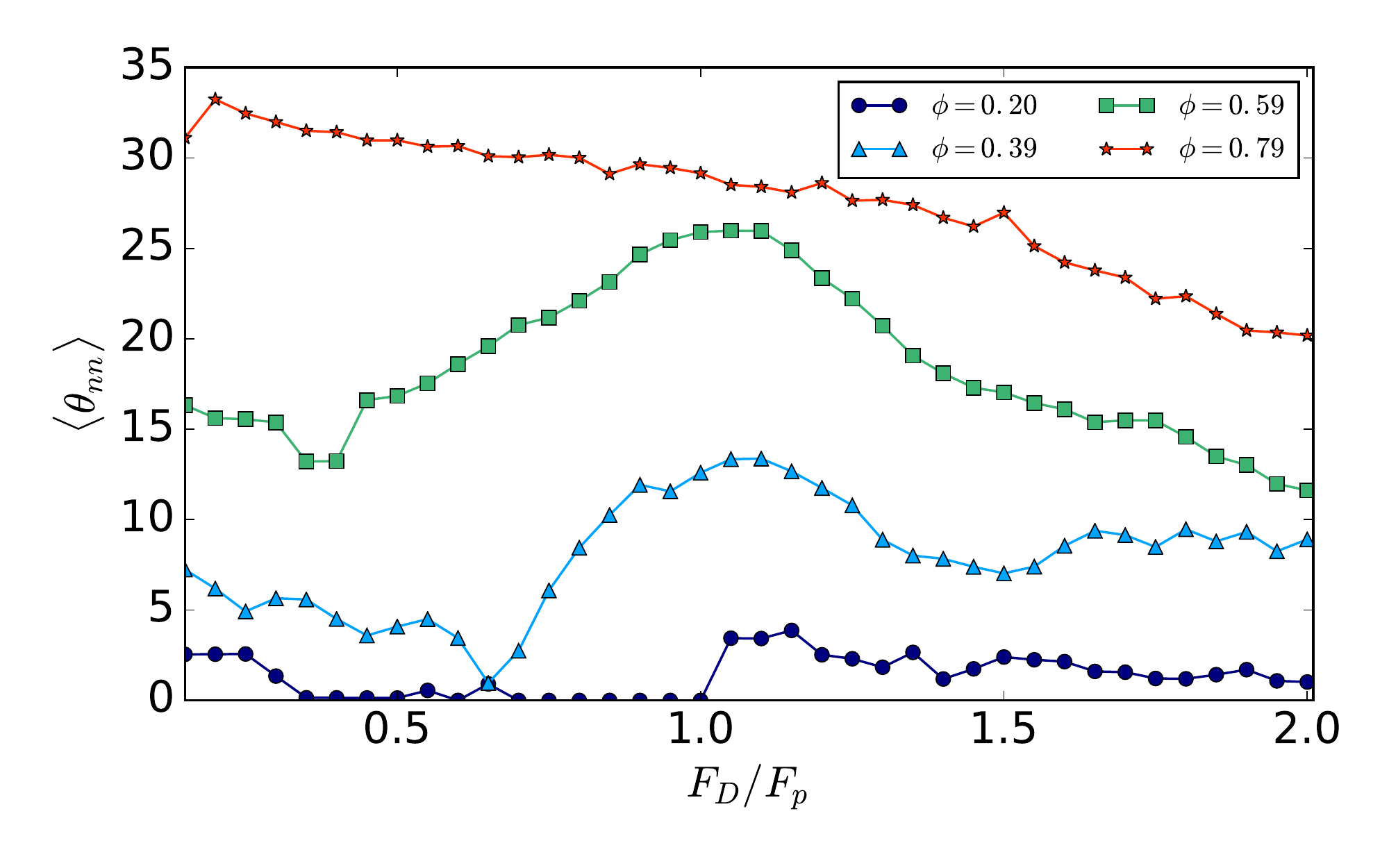}
  \caption{
    $\langle \theta_{nn} \rangle$ vs $F_D/F_p$ for the system
    in Fig.~\ref{fig:10} with $\Psi=2.0$ for
    $\phi = $ $0.20$ (circles), $0.39$ (triangles),
    $0.59$ (squares), and $0.79$ (stars).
  }
\label{fig:13}
\end{figure}

In Fig.~\ref{fig:13},
we characterize the lane structure of the disks
based on the average angle between disks that are in contact,
\begin{equation}
  \langle \theta_{nn} \rangle = \frac{1}{N_d} \sum_i^{N_d}\Theta(r^{ij}_{dd}-R_{ij}) \tan^{-1}\left(\left|\frac{{\bf R}_{ij}\cdot {\bf \hat y}}{{\bf R}_{ij} \cdot {\bf \hat x}}\right|\right) ,
\end{equation}
sampled every $\Delta t = 5 \times 10^5$ simulation time steps
after the system has reached a steady state. 
This measure is closely related to $\langle \ell_{nn} \rangle$
from Fig.~\ref{fig:8}.
Figure~\ref{fig:13} shows $\langle \theta_{nn}\rangle$ versus $F_D/F_p$ for
systems with $\phi=0.2$, 0.39, 0.59, and 0.79.
For $\phi=0.20$,
$\langle \theta_{nn} \rangle$
is low for all drives due to
the smectic structure which favors disk-disk contacts that
are aligned with the $x$ direction.
We find $\langle \theta_{nn} \rangle \approx 30^{\circ}$
near depinning
for
$\phi=0.79$,
since the polycrystalline disk arrangements
tend to contain crystallites aligned with the $x$ axis that
contribute angles of 0$^\circ$ and $60^{\circ}$ equally to the sum.
As the driving force increases,
$\langle \theta_{nn}\rangle$
decreases monotonically
due to an increase in the amount of
smectic or chainlike ordering in the system.
For $\phi=0.39$ and $\phi=0.59$,
a local maximum in $\langle \theta_{nn}\rangle$
at $F_D/F_p = 1$
is produced by the denser structures that form when the
phase separation is maximized for nearly equal
pinning and driving strengths.
This is followed by a decrease in $\langle \theta_{nn}\rangle$ at higher
drives as smectic ordering emerges.

\begin{figure}
 \includegraphics[width=0.45\textwidth]{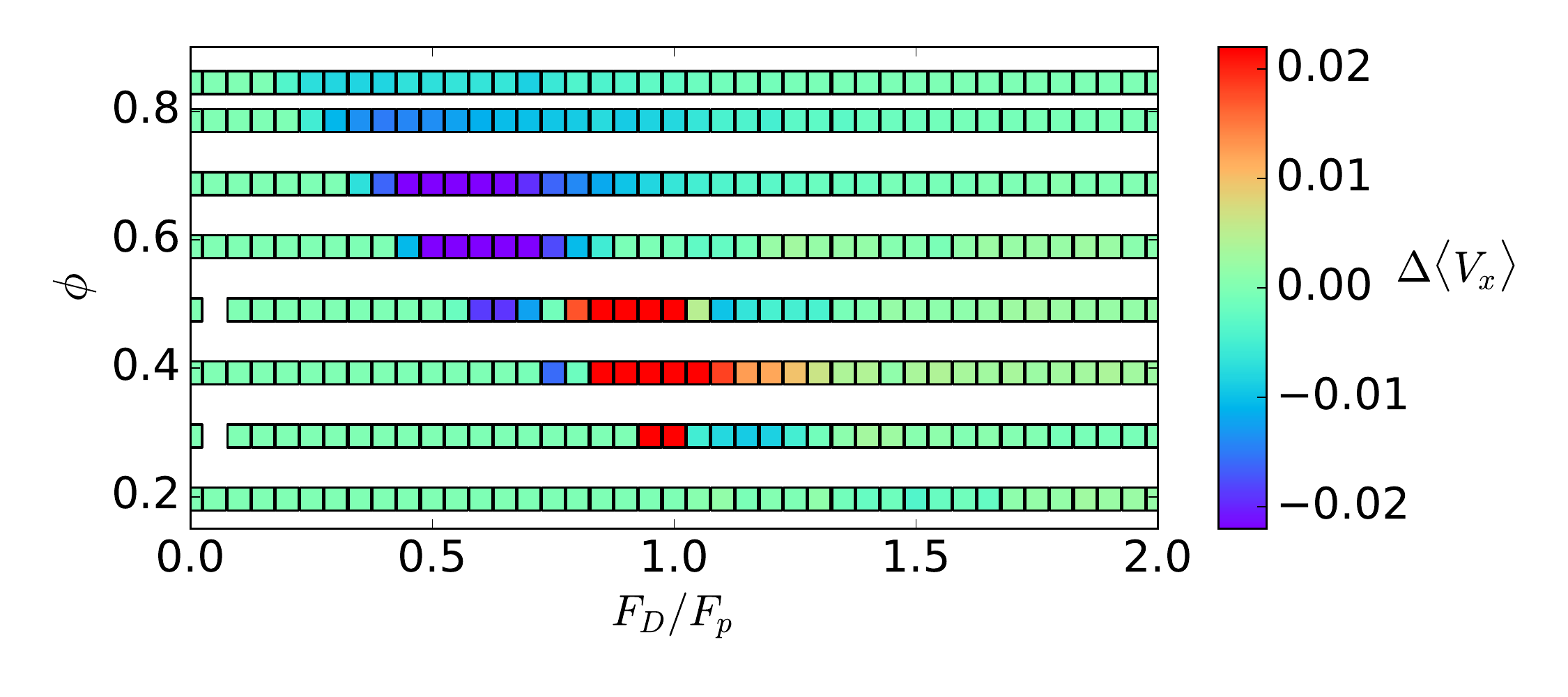}
 \caption{
   Heightfield plot of $\Delta \langle V_x\rangle$ as a function of
   total disk density $\phi$ vs driving force $F_D/F_p$, based on the
   data in Fig.~\ref{fig:10}(c).
   Red (blue) indicates that the velocity of the small disks
   is higher (lower) than that of the large disks.
   We find a large region in which
   $\Delta \langle V_x\rangle < 0$.
  }
\label{fig:14}
\end{figure}

In Fig.~\ref{fig:14}, we show a heightfield plot of $\Delta \langle V_x\rangle$
as a function of $\phi$ versus $F_D/F_p$
for the $\Psi = 2.0$ system.
Compared to the $\Psi=1.4$ system in Fig.~\ref{fig:9},
we find
a much larger region in which $\Delta \langle V_x \rangle < 0$.
This indicates that increasing the relative size of the large disks
can also increase their velocity relative to the small disks when
the driving force is close to the depinning threshold and the total
disk density is sufficiently large.

\section{Lower fraction of large disks, $N_l=N_d/10$}
\label{sec:3}
\begin{figure}
\includegraphics[width=3.5in]{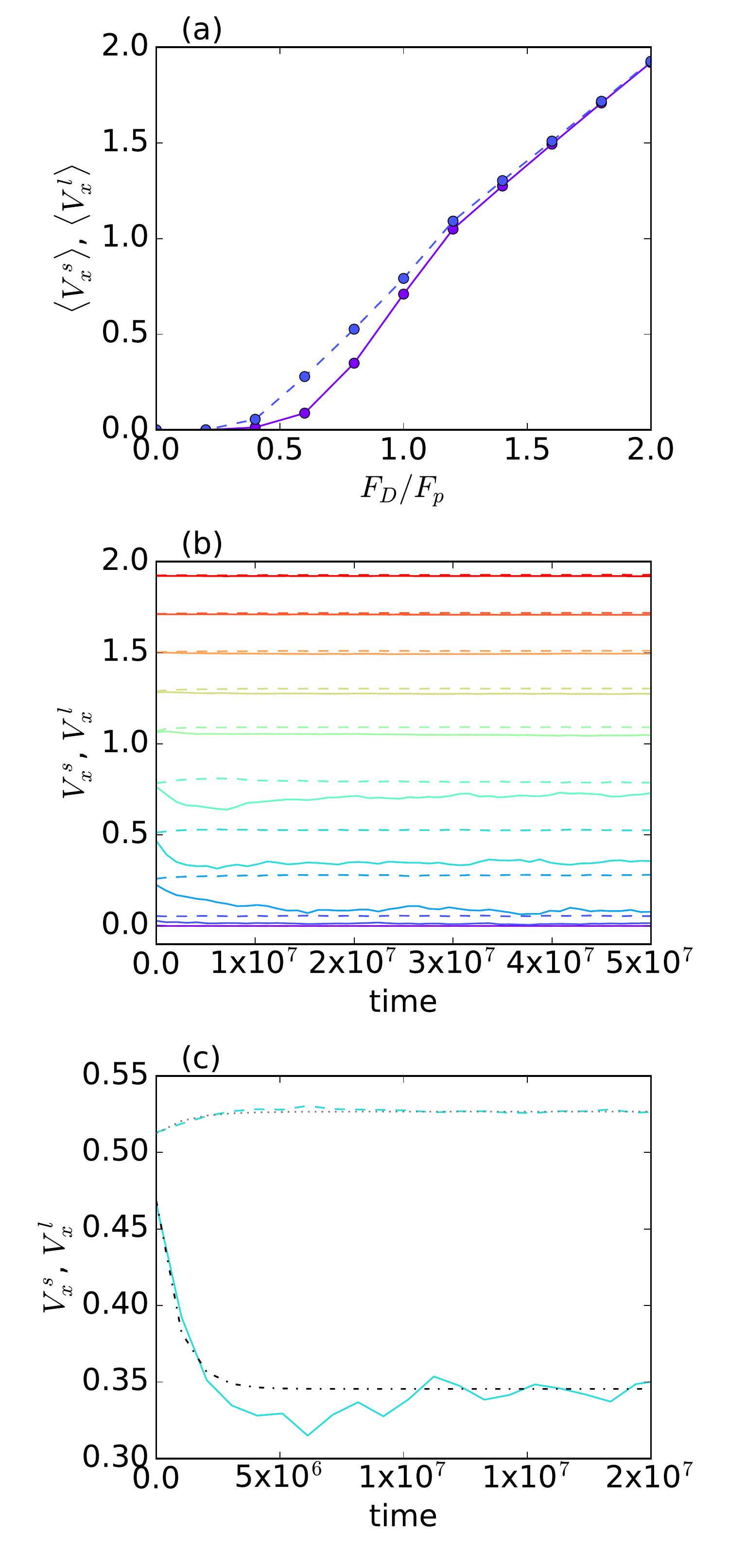}
\caption{
  (a) $\langle V_{x}^s\rangle$ (solid lines)
  and $\langle V_x^l\rangle$ (dashed lines)
  vs $F_{D}/F_{p}$
  in a sample with $\Psi=1.4$
  $N_s=0.9N_d$, and $N_l=0.1N_d$ at
  $\phi=0.48$.
  (b) The instantaneous disk velocity $V_x^s$ (solid lines)
  and $V_x^l$ (dashed lines) versus time for the small and large
  disks, respectively, in the sample from panel (a) at
  $F_D=0.2$, 
  0.4, 
  0.6, 
  0.8, 
  1.0, 
  1.2, 
  1.4, 
  1.6, 
  1.8, 
  and 2.0, 
  from bottom to top.
  The large disks reach a steady state quickly,
  while the small disks
  continue to evolve at
  $t>10^7$ timesteps.
  (c) A detail showing only the $F_D=0.8$ curves from panel (b).
  Dot-dashed line: A fit to $\langle V_x^s\rangle=e^{t/\tau_s}$ with $\tau_s=8.46 \times 10^5$.
  Dotted line: A fit to $\langle V_x^l\rangle=e^{t/\tau_l}$ with $\tau_l=1.19 \times 10^6$.
}
\label{fig:15}
\end{figure}

\begin{figure}
\includegraphics[width=0.48\textwidth]{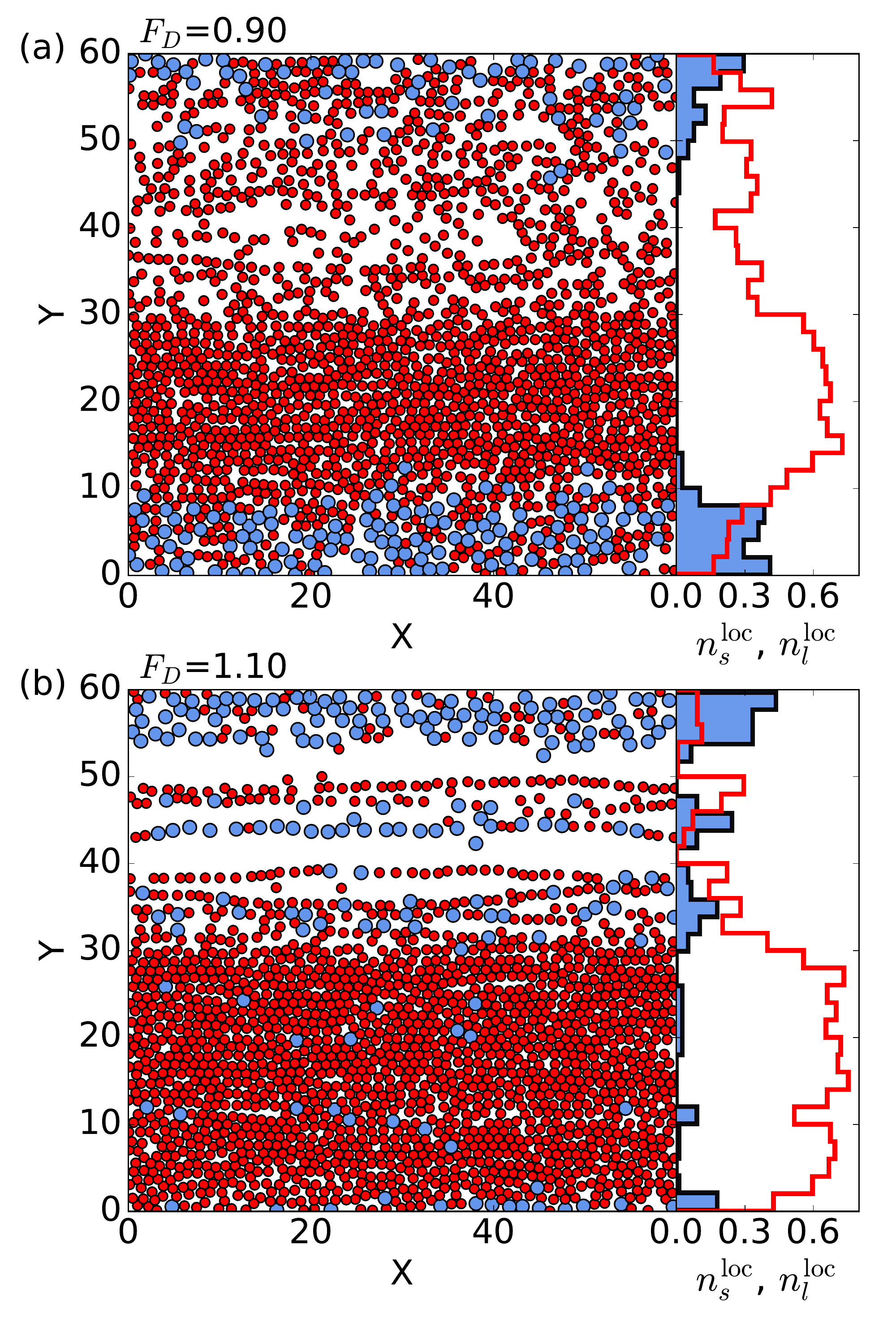}
\caption{
  Left panels: Large disk (blue circles) and small disk (red circles) positions for the
  system in Fig.~\ref{fig:15} with $\Psi=1.4$ and $N_l=0.1N_d$ at
  $\phi=0.48$.
  Right panels: $n_l^{\rm loc}$ (blue) and $n_s^{\rm loc}$ (red) as a function of $y$
  position.
  (a) $F_D/F_p=0.9$.
  (b) $F_D/F_p=1.1$.
}
\label{fig:16}
\end{figure}

\begin{figure}
  \includegraphics[width=0.45\textwidth]{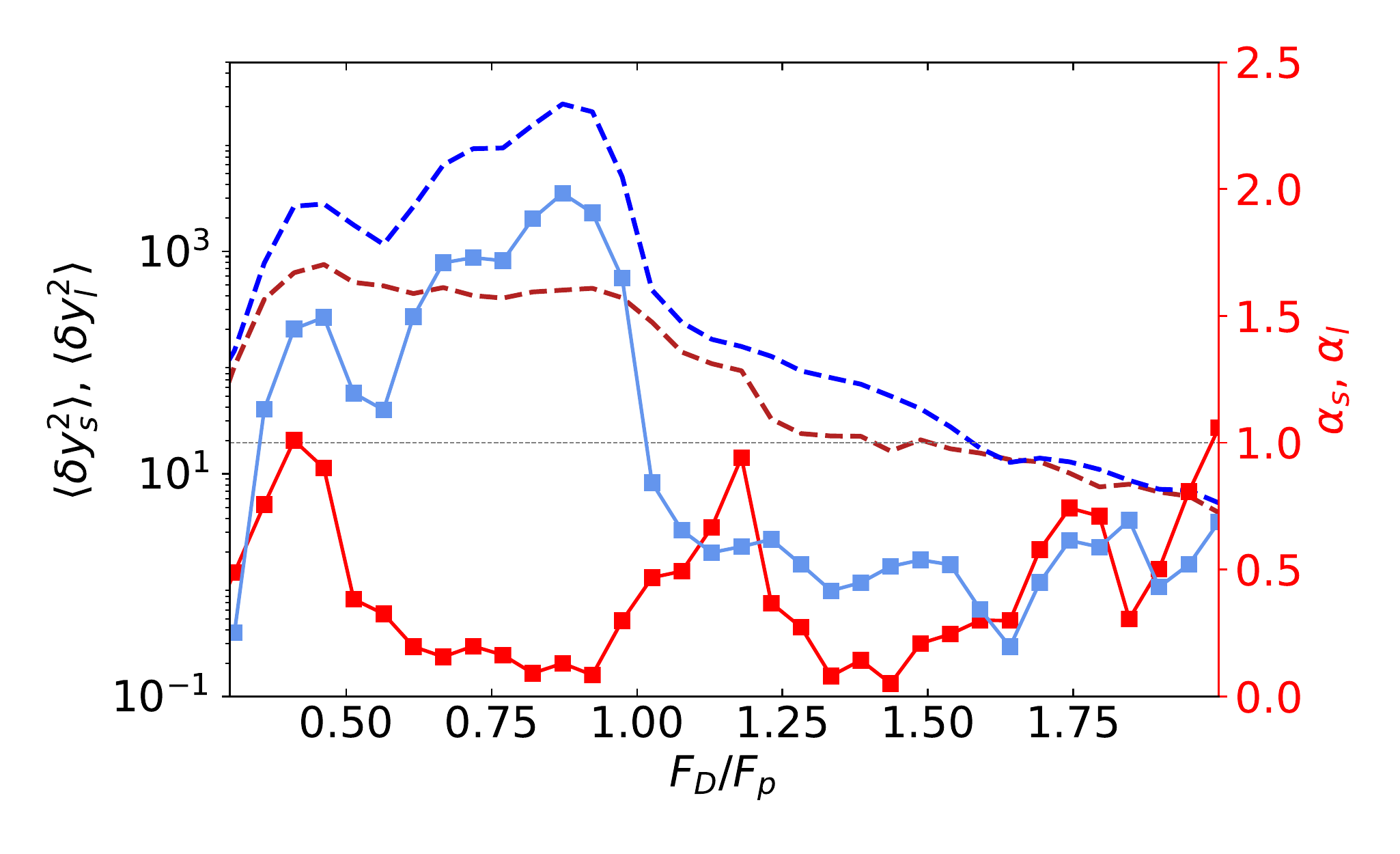}
  \caption{
    Transverse displacements
    $\langle \delta y_s^2\rangle$ (red dashed line)
    and $\langle \delta y_l^2\rangle$ (blue dashed line)
    for the small and large disks obtained after $1 \times 10^7$ simulation
    time steps vs $F_D/F_p$ and the corresponding diffusive
    exponent $\alpha_s$ (red squares) and $\alpha_l$ (blue squares) for the system
    in Fig.~\ref{fig:15} with $\Psi=1.4$ and $N_l=0.1N_d$ at $\phi=0.48$.
}
\label{fig:17}
\end{figure}

We next investigate the effect of changing the disk species ratio
from $N_s=N_l=0.5N_d$
to $N_s=0.9N_d$ and $N_l=0.1N_d$ for a system with
$\Psi=1.4$.
We find the same general phases as described in Sec. III but
with
a greater tendency for the large disks to move faster than
the small disks.
In Fig.~\ref{fig:15}(a), 
we
plot $\langle V_x^s\rangle$ and $\langle V_x^l\rangle$ versus $F_D/F_p$
for the $N_l=0.1N_d$ system at
a disk density of $\phi=0.48$.
We find plastic depinning for both disk species, as indicated by the
concave shape of the velocity-force curve, followed by a
transition at higher drives to a linear dependence.
At $F_D/F_p=0.9$, illustrated in Fig.~\ref{fig:16}(a), the system can be
divided into three regions:
a small disk liquid, a small disk gas, and a mixed gas-like region containing
both disk species at an intermediate density.
At a higher drive of $F_D/F_p=1.1$ in Fig.~\ref{fig:16}(b),
the small disk liquid has increased in density and contains a few large disks.
A window of large disk liquid containing some small disks runs along one side of
the small disk liquid, while
the low density region of the sample contains roughly
equal numbers of small and
large disks arranged in a smectic structure.
Due to the strong species segregation,
these phases resemble the states found for monodisperse disks
in Ref.~\cite{22}.
Over the range
$0.2 < F_D/F_p < 1.6$ where the species separation occurs,
$\langle V_{x}^l \rangle > \langle V_{x}^s \rangle$,
giving $\Delta \langle V_x\rangle < 0$ (not shown).

In Fig.~\ref{fig:15}(b), 
we plot the time evolution of $\langle V_x^s \rangle$
and $\langle V_x^l\rangle$
for the
same
$\phi = 0.48$ system at $F_D$ values ranging from $F_D/F_p=0.2$ to
$F_D/F_p=2.0$.
For $F_D/F_p \leq 0.2$,
the system is pinned and
$\langle V_x^s \rangle = \langle V_x^l\rangle =0$.
When $F_D/F_p \ge 0.4$,  
we find $\langle V_{x}^l \rangle > \langle V_{x}^s \rangle$,
with
$\langle V_x^l\rangle$ remaining nearly constant over time while
$\langle V_x^s\rangle$ decays.
For
intermediate driving forces of $0.6 < F_D/F_p < 1.2$,
the $\langle V_x^s\rangle$ curves have an exponential shape,
$\langle V_x^s \rangle \propto e^{-t/\tau_s} + \langle V_o \rangle$,
as shown
in Fig.~\ref{fig:15}(c)
for $F_D / F_p = 0.8$,
where
$\tau_s = 8.46 \times 10^5$.
A similar fit of $\langle V_x^l\rangle$ at the same drive
gives a time constant
$\tau_l = 1.19 \times 10^6$ that is
somewhat larger.
As $F_D/F_p$ increases above $1.2$,
the system
rapidly reaches
a steady state and the difference between the velocity of the
small and large disks vanishes.
Due to the lengthy transient dynamics at intermediate
$F_D/F_p$,
we wait a minimum of $2 \times 10^7$ simulation time steps
before measuring the velocity-force curves shown
in Fig.~\ref{fig:15}(a).

In Fig.~\ref{fig:17} we plot the transverse displacements
$\langle \delta y_s^2 \rangle$ and $\langle \delta y_l^2\rangle$ versus
$F_D/F_p$ for the $\phi=0.48$ sample along with the
corresponding exponents $\alpha_s$ and $\alpha_l$.
All four quantities increase monotonically between
$F_D = F_c$
and $F_D/F_p = 0.4$.
At intermediate $F_D$,
we find subdiffusive transverse motion of the
small disks with 
$\alpha_s<1$
accompanied by superdiffusive transverse motion of the
large disks with $\alpha_l>1$.
Here the small disks are
confined within a dense liquid,
while the large disks are in a low density region in which
interactions with pinning sites can enhance
the transverse diffusion.
At large $F_D$ where smectic structures emerge, both disk species
have subdiffusive transverse motion.

\section{Scaling Near The Depinning Transition}
\label{sec:4}
In systems of particles that have long range interactions,
the velocity-force relationship scales
as
$V \propto (F_D - F_c)^{-\beta}$ \quad \cite{1}.
For elastic depinning in which the
structure of the particle lattice remains unchanged,
$\beta=2/3$, while
when the depinning transition is plastic,
$\beta > 1.0$.
For
Coulomb \cite{6} and screened Coulomb \cite{14,33} interaction potentials,
the plastic depinning 
exponents are $\beta \approx 1.65$ and $2.0$,
respectively, while
simulations of depinning of
superconducting vortices with a Bessel function vortex-vortex
interaction give $\beta = 1.3$\cite{33}.
It is interesting
to ask whether similar scaling of the velocity-force curves
occurs in the disk system.
For monodisperse disks with
$N_p / N_d > 0.288$,
it was shown in Ref.~\cite{22} that
the velocity-force curves can be fit to a power law with
$1.4 < \beta < 1.7$.

\begin{figure}
\includegraphics[width=0.46\textwidth]{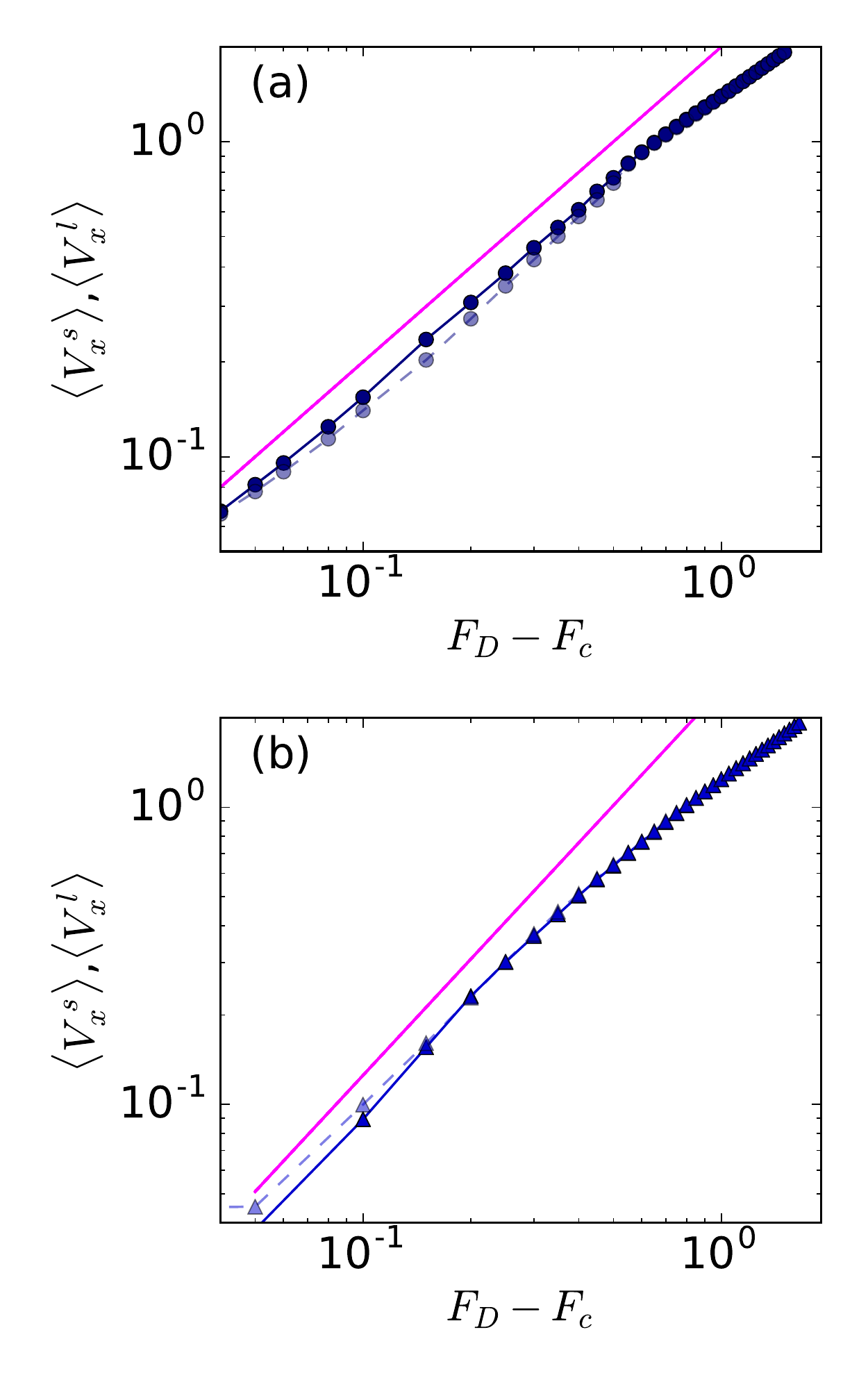}
\caption{
  $\langle V_x^s\rangle$ (solid lines) and $\langle V_x^l\rangle$ (dashed lines)
  vs $F_D-F_c$ on a log-log scale
  for the sample from Fig.~\ref{fig:1} with $\Psi=1.4$ and $N_s=N_l$.
  We fit the data to $\langle V_x^{s(l)}\rangle\propto (F_D-F_c)^{-\beta}$ (pink lines).
  (a) $\phi=0.46$ with $\beta=1.0$.
  (b) $\phi=0.58$ with $\beta=1.3$.
}
\label{fig:18}
\end{figure}

In Fig.~\ref{fig:18}(a,b) 
we plot $\langle V_x^s \rangle$  and $\langle V_x^l\rangle$ versus
$F_D - F_c$ on a log-log scale
at densities of $\phi=0.46$ and 0.58, respectively.
By fitting the portion of the curve closest to depinning, we find
$1.0 < \beta < 1.3$.
The scaling fit can be performed only for $\phi > 0.35$ and does not work
at low disk densities.  We find similar scaling fits for sufficiently large
disk densities for the $\Psi=2.0$ system and for the $\Psi=1.4$ and $N_l=0.1N_d$
system.
The depinning is clearly not elastic, but the lower values of $\beta$ compared to
systems with longer range interactions suggest that the type of plastic depinning
that occurs may be different for short range interacting systems than for longer range
interacting systems.

\section{Summary}
\label{sec:5}

We examine the dynamics of
bidisperse disks
driven over random quenched disorder to explore
the dynamical phases of particles
with short range interaction forces. 
At low disk densities,
we observe a pinned state that transitions
into a strongly
chained state
where the disks can undergo local demixing but
where the overall disk distribution is homogeneous.
At intermediate disk densities,
the disks depin into
a disordered flow state
exhibiting stick slip dynamics, followed
by a species segregated state in which the
small disks form clusters and the large disks remain evenly
distributed throughout the sample.
For intermediate drives the disks form
a partially laned state
exhibiting both species separation and density segregation, while at high
drives a mixed laning state emerges.
At high disk densities of $\phi > 0.75$,
a rigid polycrystalline state appears that
moves as a solid and undergoes
no species or density segregation.
Both the density and the species
segregation effects are
the most prominent near $F_D=F_p$ when the driving force
and pinning force directly compete.
The anisotropic fluctuations induced by the pinning at high drives
favor the formation of laned states.
It is also possible to induce mixing
between the two species
just above the 
depinning transition.
By increasing the radius of the large disks
compared to that of the small disks, we find a larger amount of
crystallization and banding of the large disks, while the small disks tend
to form an interstitial liquid.
Lowering the fraction of large disks compared to the fraction
of small disks tends to increase the velocity of the large disks compared
to that of the small disks, which species separate into a disordered liquid
that flows unevenly over the pinning sites.
When the disk density is sufficiently large, we find scaling of the velocity-force
curves near the plastic depinning transition with an exponent that is slightly
smaller than what is observed in systems with longer range interparticle interactions,
suggesting that the plastic depinning transition may have distinct features when the
interaction range is very short.

Our results could be relevant to
multi-species flows of soft matter through random substrates or the 
flow of granular matter over a disordered background.
It would be interesting to explore
possible segregation effects
for bidisperse systems with long range
particle-particle interactions driven over random disorder. 
In the disk system, the segregation of particles
into clumps reduces the number of disk-disk collisions
and enhances the disk flow.

\acknowledgments
This work was carried out under the auspices of the 
NNSA of the 
U.S. DoE
at 
LANL
under Contract No.
DE-AC52-06NA25396.
This research was supported in part by
the University of Notre Dame Center for Research Computing
and the Wabash College Computational Chemistry Cluster.

\end{document}